\documentclass{aa}
\usepackage{amssymb}
\usepackage{amsmath}
\usepackage{graphicx}
\usepackage{url}
\usepackage{txfonts}
\usepackage{natbib}
\usepackage{float}
\usepackage{array}
\usepackage{epstopdf}
\usepackage{color}
\usepackage[dvipsnames]{xcolor}
\usepackage{multicol}
\usepackage{siunitx}
\usepackage{multirow}
\usepackage{hhline}
\begin{document} 

   \title{Weak lensing magnification of SpARCS galaxy clusters}
   \subtitle{}
   \author{A. Tudorica
          \inst{1},
          H. Hildebrandt\inst{1},
          M. Tewes\inst{1},
          H. Hoekstra\inst{2},
          C.B. Morrison\inst{3},
          A. Muzzin\inst{4},
          G. Wilson\inst{5},
          H.K.C. Yee\inst{6},
          C. Lidman\inst{7,8},
          A. Hicks,\inst{9},
          J. Nantais,\inst{10},
          T. Erben\inst{1},
          R. F. J. van der Burg\inst{11}
          \and
          R. Demarco\inst{12}
          }
          \authorrunning{A.Tudorica et. al.}

\institute{Argelander Institute for Astronomy (AIfA), University of Bonn, Auf dem H\"uegel 71, 53121 Bonn, Germany\\
              \email{tudorica@astro.uni-bonn.de}
   \and{Leiden Observatory, Leiden University, P.O. Box 9513 2300 RA Leiden, The Netherlands}
   \and{Department of Astronomy, University of Washington, Box 351580, Seattle, WA 98195, USA}
   \and{Department of Physics and Astronomy, York University, 4700 Keele Street, Toronto, Ontario, ON MJ3 1P3, Canada}
   \and{Department of Physics and Astronomy, University of California-Riverside, 900 University Avenue, Riverside, CA 92521, USA}
   \and{Department of Astronomy and Astrophysics, University of Toronto, 50 St. George Street, Toronto, Ontario, M5S 3H4, Canada}
   \and{School of Physics, University of Wollongong, Wollongong, NSW 2522, Australia}
   \and{Australian Astronomical Observatory, North Ryde, NSW, Australia}
   \and{Cadmus, Energy Services Division, 16 N. Carroll Street, Suite 900, Madison, WI 53703, United States}
   \and{Departamento de Ciencias F\'isicas, Universidad Andres Bello, Fernandez Concha 700, Las Condes 7591538, Santiago, Chile} 
   \and{Laboratoire AIM-Paris-Saclay, CEA/DSM-CNRS-Universit\'e Paris Diderot, Irfu/Service d’Astrophysique, CEA Saclay, Orme des Merisiers, 91191 Gif-sur-Yvette, France}
   \and{Departamento de Astronom\'ia, Universidad de Concepci\'on, Casilla 160-C, Concepci\'on, Chile}\\
\\  
             }

   \date{Received 29.05.2017; accepted 03.10.2017}
   
  \abstract
   {Measuring and calibrating relations between cluster observables is critical for  resource-limited studies. The mass-richness relation of clusters offers an observationally inexpensive way of estimating masses. Its calibration is essential for cluster and cosmological studies, especially for high-redshift clusters. Weak gravitational lensing magnification is a promising and complementary method to shear studies, that can be applied at higher redshifts.}
   {We aim to employ the weak lensing magnification method to calibrate the mass-richness relation up to a redshift of 1.4. We used the Spitzer Adaptation of the Red-Sequence Cluster Survey (SpARCS) galaxy cluster candidates ($0.2<z<1.4$) and optical data from the Canada France Hawaii Telescope (CFHT) to test whether magnification can be effectively used to constrain the mass of high-redshift clusters.}
   {Lyman-Break Galaxies (LBGs) selected using the $u$-band dropout technique and their colours were used as a background sample of sources. LBG positions were cross-correlated with the centres of the sample of SpARCS clusters to estimate the magnification signal, which was optimally-weighted using an externally-calibrated LBG luminosity function. The signal was measured for cluster sub-samples, binned in both redshift and richness.}
   {We measured the cross-correlation between the positions of galaxy cluster candidates and LBGs and detected a weak lensing magnification signal for all bins at a detection significance of 2.6-5.5$\sigma$. In particular, the significance of the measurement for clusters with $z>1.0$ is 4.1$\sigma$; for the entire cluster sample we obtained an average M$_{200}$ of 1.28 $^{+0.23}_{-0.21} \times 10^{14} \, \textrm{M}_{\sun}$.
  }
   {Our measurements demonstrated the feasibility of using weak lensing magnification as a viable tool for determining the average halo masses for samples of high redshift galaxy clusters. The results also established the success of using galaxy over-densities to select massive clusters at $z > 1$. Additional studies are necessary for further modelling of the various systematic effects we discussed.}

               
   \maketitle

\section{Introduction}

The statistical properties of the distribution of mass in the Universe is one of the fundamental predictions of any cosmological model. The properties of the large scale structure can be used as powerful constraints on cosmological parameters. Galaxy clusters are the largest gravitationally bound structures in the Universe, containing hundreds to thousands of galaxies, with typical masses ranging between $10^{14}$-$10^{15}$\,M$_\sun$  \citep[see][for reviews]{Voit2005,Allen2011}. Besides galaxies, clusters also contain large amounts of dark matter and hot, X-ray emitting intra-cluster gas. One very useful property of galaxy clusters is that the relative proportions between these three main components remain approximately constant, therefore the total mass can be estimated by measuring the properties of only one component, resulting in several scaling relations \citep[see][for a comprehensive review]{Giodini2013}. The correlation between the total mass of a galaxy cluster and the number of galaxies belonging to it (richness) is one of the most accessible scaling relations. However, clusters evolve with time and interact with each other; consequently the scaling relations must be calibrated to consider these changes. At high redshift ($z>0.8$), galaxy clusters are observed while they are still in the assembly phase and therefore, certain assumptions about their dynamical state required for an accurate estimate using the X-ray emissions or velocity dispersions of galaxies, might not hold any more.

Gravitational lensing has the unique property among different methods of mass measurement that it is sensitive to all of the mass along the line of sight, not differentiating between dark and baryonic matter and being independent of assumptions regarding the dynamical state of matter \citep[for an in-depth review, see][]{Bartelmann2001}. There are various ways of using the gravitational deflection of light to determine the properties of massive objects, each with its own set of advantages and disadvantages. Multiple images and strong lensing arcs are most useful for studying the innermost areas of galaxy clusters and obtaining precise mass estimates, but these methods are applicable only in the case of very massive clusters. In contrast, weak lensing shear measurements are based on the statistical properties of the minute deformations measured for the observed shape of background galaxies. Modelling of the shear distortion has been intensively studied, developing into a set of reliable methods of measurement for stacks of cluster samples, and even for individual clusters which are situated at the high end of the mass spectrum \citep{Gruen2014,Umetsu2014,Applegate2014,vonderLinden2014a,vonderLinden2014b,Hoekstra2015}. Shear-based weak lensing techniques can be applied to a wide range of clusters, but since these measurements rely on precise measurements of shapes, this requires galaxies to be resolved. This limits the applicability of the method to low redshifts using ground-based telescopes. Space-based telescopes such as the Hubble Space Telescope provide an alternative to this issue, albeit an observationally expensive one \citep{Schrabback2017}. 

We employed a third method based on gravitational lensing to estimate the cluster masses: the weak lensing magnification effect. Magnification is a geometric consequence of gravitational lensing, equivalent to an enlargement of the observed solid angle. The sources appear to have a greater angular size and because the surface brightness remains constant, the observed flux will be amplified accordingly. It has the advantage of providing mass estimates at higher redshifts compared to other lensing methods, while using less demanding observational data. Although the signal-to-noise ratio (S/N) provided by magnification measurements is lower per galaxy than the corresponding shear-based estimate, this is partly compensated by the fact that magnification does not require the use of resolved sources unlike shear-based methods. The magnification component of lensing has been measured with increasing accuracy and precision in recent years, with some studies taking advantage of the complementarity between shear and magnification for cluster analyses \citep[e.g.][]{Umetsu2011a,Umetsu2014}.

This work used Lyman Break Galaxies (LBGs) as background sources \citep[see][]{Steidel1996,Giavalisco2002}, measuring the magnification-induced deviation of the source number densities from the average \citep[several studies have previously used this method, see][]{Hildebrandt2009, Hildebrandt2011, Hildebrandt2013, Morrison2012, Ford2012, Ford2014}. Using LBGs as background sources brings several important advantages, such as the fact that their redshift distribution is well known, contamination at low redshift is small and the spatial density is higher than for other sources previously used in studies using similar techniques (e.g. quasars). 

For any magnitude bin, the observed number density of sources can increase or decrease, depending on the local slope of the luminosity function \citep[i.e. source magnitude, see][]{Narayan1989}. Through stacking the signal from a sample of clusters, the signal-to-noise ratio of the measurements is boosted, while the dominant source of noise, the physical clustering of the background (source) galaxies, is averaged out.

The  magnification signal was measured by using cross-correlation between cluster centres and LBG candidate positions. This method provides an estimate of the average $M_{200}$ for the sample of clusters used for measuring the cross-correlation. The magnification signal is modelled with a composite large scale structure halo model, taking cluster miscentering and low-redshift contamination of sources into account.

In Sect. \ref{data} we described the data. Optical data reduction, selection procedures for cluster candidates, sources and systematic tests are discussed in Sect. \ref{data_reduction}. In Sect. \ref{methods} we detailed the methodology for measuring and modelling of the magnification signal. The results were presented in Sect. \ref{results} and discussed in Sect. \ref{Section:Discussion}, while the conclusions can be found in Sect. \ref{conclusions}. The cosmological model used in this paper is based on the standard Lambda Cold Dark Matter ($\Lambda$CDM) cosmology with $H_0=67.3$\,km\,s$^{-1}$\,Mpc$^{-1}$, $\Omega_M=0.316$, $\Omega_{\Lambda}=1-\Omega_M=0.684$ and $\sigma_8=0.83$ \citep[see][]{Planck2015}, while distances are in megaparsecs. All magnitudes throughout the paper are in the AB system.

\section{Data}\label{data} 
\subsection{Infrared}

\begin{figure}[htb]
 \centering
\includegraphics[width=91mm]{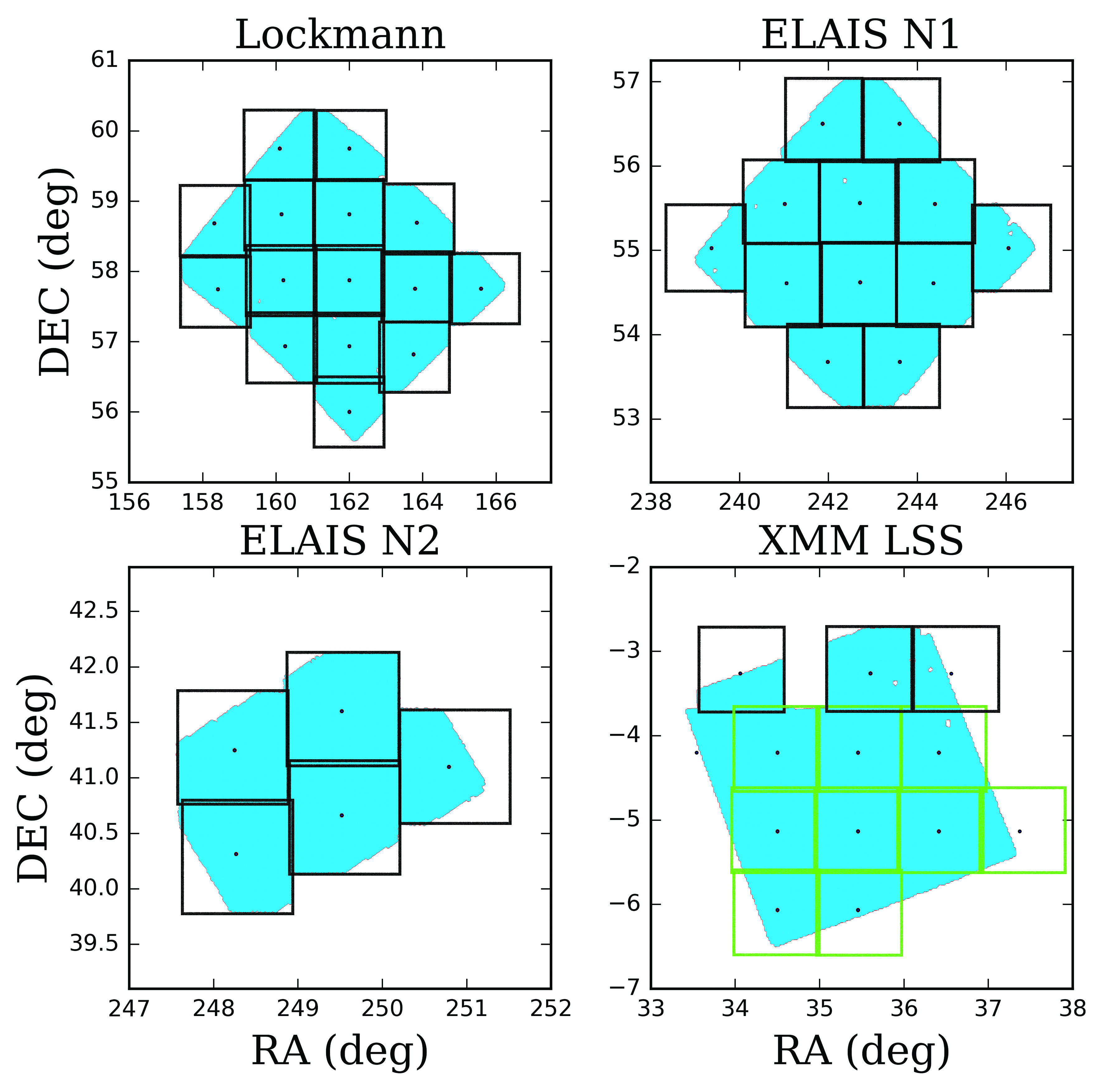}
 \caption{Outline of the SpARCS fields observable from the northern hemisphere. The blue area traces the distribution of sources detected in SWIRE (with the original data  masking applied), while the black squares show the locations of the CFHT individual pointings, each covering approximately 1 deg$^2$. The bottom green squares in the XMM\,LSS field outline the CFHTLS pointings we use. Pointing centres are marked with black dots.}
 \label{fig:north}
\end{figure}

The Spitzer Wide-area InfraRed Extragalactic Legacy Survey\footnote{http://swire.ipac.caltech.edu/swire/public/survey.html} (SWIRE) \citep[][]{Lonsdale2003} is one of the six large legacy surveys observed during the first year in space of the Spitzer Space Telescope. It covers approximately 50 deg$^2$ in all 7 infrared wavelength bands available on Spitzer: four with the Infrared Array Camera \citep[IRAC, see][]{Fazio2004} at 3.6, 4.5, 5.8, 8\,$\mu$m, and three more with the Multiband Imaging Photometer for Spitzer \citep[MIPS, see][]{Rieke2004}, at 24, 70 and 160\,$\mu$m. The survey is divided in six separate patches on the sky, with three located in the northern hemisphere (European Large Area ISO Survey - ELAIS - N1, N2 and the Lockman Hole), two of the fields in the southern hemisphere (ELAIS S1 and Chandra Deep Field South) and one equatorial field, XMM-Newton Large Scale Structure Survey (XMM LSS). We used only the XMM LSS, ELAIS N1\&N2 and the Lockman hole fields in this study. Figure \ref{fig:north} shows the outline of the four SWIRE fields that overlap with the CFHT data and the individual CFHT pointings used in this work.
\begin{table*}[htb]\setlength\extrarowheight{2.5pt}   
\caption{Properties of the four SpARCS fields used in this study.}
\centering
\begin{tabular}{ccccccc}
\hline
\hline
Field 	Name		&	RA	(centre) &	Dec (centre)	& SWIRE 3.6$\,\mu$m Area&	SpARCS Area		&  Usable Overlap Area  & Passbands\\
              		&   $\left(\,\textrm{HH:MM:SS}\right)$	& $\left(\,\textrm{DD:MM:SS}\right)$		&        $\left(\,\textrm{deg}^2\right)$   	&  $\left(\,\textrm{deg}^2\right)$		&  $\left(\,\textrm{deg}^2\right)$		& 				  \\
\hline                    
 XMM LSS			&	02:21:20	&	-04:30:00		&	9.4				&	11.7		&	7.3			& $u g r i z$	  \\
 Lockman Hole		&	10:45:00	&	+58:00:00		&	11.6			&	12.9		&	9.7			&	$u g r z $	  \\
 ELAIS N1			&	16:11:00	&	+55:00:00		&	9.8				&	10.3		&	4.3			&	$u g r z$     \\
 ELAIS N2			&	16:36:48	&	+41:01:45		&	4.4				&	4.3			&	3.4			&	 $u g r z$    \\
 \hline
 Total			&				&					& 	50.4			&	55.4		&	41.9		&	         
\end{tabular}
\label{SpARCS}
\end{table*}   

\subsection{Cluster catalogue}

The Spitzer Adaptation of the Red-Sequence Cluster Survey (SpARCS, \cite{Wilson2009,Muzzin2009}) is a follow-up survey of the SWIRE fields in the $z'$ band down to a mean depth of $z_{AB}'=24.0$ at 5$\sigma$ (for extended sources), using MegaCam on the 3.6\,m CFHT for the three Northern fields and XMM LSS, while  MOSAIC II  was used on the 4\,m Blanco telescope at the Cerro Tololo Inter-American Observatory (CTIO) for the Southern Fields. It is one of the largest high-z cluster surveys with a total area of $41.9$\,deg$^2$, with hundreds of $z>1$ cluster candidates based on the $z'-3.6$\,$\mu$m colour.
 
The SpARCS cluster catalogue was created by using a modified version of the \cite{Gladders2000} algorithm, as described in detail by \cite{Muzzin2008}.  The Cluster Red-Sequence (CRS) method employed requires the use of only two imaging passbands that span the rest-frame of the \SI{4000}{\angstrom} break feature in early type galaxies. Elliptical galaxies constitute the dominant population in galaxy clusters, lying along a linear relation in colour-magnitude space. In the colour-magnitude diagram constructed with such a combination of filters, elliptical galaxies in clusters appear always as the reddest and brightest at any specific redshift, strongly contrasting with the field population. 

\begin{figure}[ht]
\centering
\includegraphics[width=92mm]{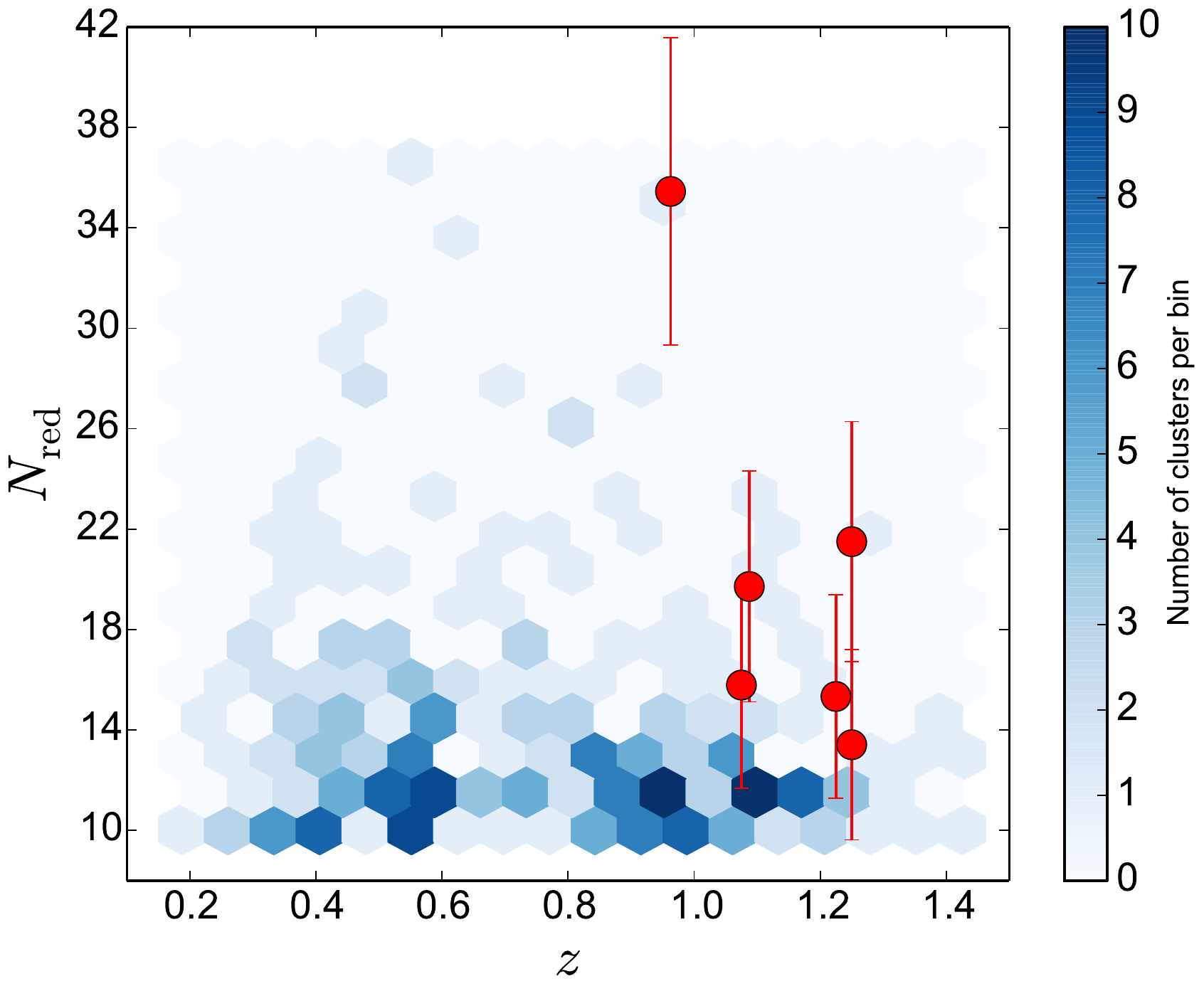}
\caption{Number density as a function of redshift and richness for the sample of galaxy clusters used in this study. The six GCLASS clusters falling within the area covered by the CFHT data are shown individually with the red points, with the errorbars representing the  uncertainty in their $N_{\mathrm{red}}$ values.}
\label{fig:cluster_hist}
\end{figure} 

\citet{Muzzin2009} and \citet{Wilson2009} construct the cluster candidates catalogue by finding peaks in the smoothed spatial galaxy density maps of individual colour slices representing different redshifts. Galaxies are given weights based on several criteria. In addition to weights based on their colours, galaxies are also weighted based on their apparent magnitude, relative to a fiducial M* value, since early type cluster galaxies are usually the reddest and brightest galaxies within a colour slice. The probability of belonging to a colour sequence model line for a particular galaxy is also taken into account by weighting. A probability map is constructed for each colour slice by considering the aforementioned weights, representing the spatial galaxy density map of the survey within each redshift slice. The pixel size for each map is 125kpc at all redshifts. The galaxies within each pixel are added, weighted by the product of the corresponding colour and magnitude weights. Each map has the noise properties homogenized by smoothing with an exponential kernel and by adding redshift dependent noise maps. We refer the reader to Sec. 3.1-3.6 of \cite{Muzzin2008} for a detailed description of the cluster detection algorithm and to \cite{Muzzin2009} and \cite{Wilson2009} for more details on its application to the SpARCS dataset.

The richness parameter associated with these detections is quantified by $N_{\mathrm{red}}$, a slightly altered version of the cluster-center galaxy correlation amplitude ($B_{\textrm{gc}}$) estimator described in detail by \cite{Yee1999}. $N_{\mathrm{red}}$ represents the number of background-subtracted, red-sequence galaxies brighter than $M^*+1$ within a 500 kpc circular aperture. $M^*$ is determined from the survey data \citep[see Sec. 5.1 and Fig. 14 in][]{Muzzin2008}, while the width of the red-sequence is chosen to be $\pm 0.15$ mag at all redshifts. The scaled version of $N_{\mathrm{red}}$, $B_{\textrm{gc}}$ has been shown to correlate well with various cluster properties \citep[e.g. $R_{200}$, X-ray temperature, velocity dispersion, virial radius, see][]{Yee1999,Yee2003,Gilbank2004,Muzzin2007b}.

The exact position of the cluster centre is a critical piece of information as many important properties are estimated using measurements that depend significantly on the approximated centre position (e.g. richness, mass, luminosity function etc.). \cite{Muzzin2008} estimate two centroids, one based on the location of the peak of the red sequence probability flux in the probability maps, and the other defined as the position where the $N_{\mathrm{red}}$ is maximized. We correct for cluster miscentering statistically in the model by shifting the cluster centers with a radial offset following a 2-D Gaussian probability distribution \citep[see Fig. 1 in][]{Ford2014}. Since the difference between these two centroid estimates is small and it does not make a significant difference in the final results, we chose to use only the former cluster centre estimates from \cite{Muzzin2008}, that is the position where the $N_{\mathrm{red}}$ is maximised.

The CRS technique is well tested and is an observationally efficient method for selecting galaxy clusters in high-redshift surveys \citep{Gladders2005,Wilson2005b}, providing photometric redshifts accurate to 5 percent \citep{Gilbank2007,Blindert2004} as well as low false-positive rates (smaller than 5\%, see for example \cite{Gilbank2007,Blindert2004,Gladders2005}). As part of the Gemini CLuster Astrophysics Spectroscopic Survey (GCLASS), 10 of the richest cluster candidates in SpARCS with a photometric redshift range $0.86\leq z\leq 1.34$ were observed spectroscopically over 25 nights with the Gemini North and South telescopes, confirming their cluster nature and their distance estimated with the CRS algorithm \citep{Muzzin2012,vanderBurg2014}.
 
We selected 287 candidate clusters from the SpARCS catalogue compiled by \citet{Muzzin2009} and \citet{Wilson2009}, with a cut-off in richness $N_{\mathrm{red}} \geq 10$, which ensures that the detection significance is high and the candidate has a high likelihood of being a real galaxy cluster. The distribution of redshifts and of the $N_{\mathrm{red}}$ richness for the sample, along with six individual clusters from GCLASS can be seen in Fig. \ref{fig:cluster_hist}. 

\subsection{Optical $ugriz$}
 
We added $ugri$ coverage to the Northern SpARCS fields from available CFHT archival and proprietary data, with the total area and available filters for each patch described in Table \ref{SpARCS}. The MegaCam instrument is mounted in the CFHT prime focus and consists of 36 charge-coupled devices ($2048 \times 4612$\,pixels each, totalling 340\,megapixels) with a pixel scale of $0\farcs187$ and covering a total field of view of about 1\,deg$^2$. 

We obtained 35 individual CFHT MegaCam pointings designed to maximise the total overlap with the SWIRE fields. Coverage in the $i$-band is available only for the pointings overlapping the XMM LSS area\footnote{The corresponding CFHT proposal identification codes (PIDs) for the SpARCS optical data are: 12AC02, 12AC99, 12BC05, 11BC97 and 11BC23.}. We aimed to have a uniform depth for the fields in all bands, complementing existing data with our observations. The $r$-band average depth goal was $r \lesssim 24.5$, since the brightest LBGs ($\lesssim 24.5$) carry the largest signal. Table \ref{Table:data_properties} contains the average seeing, limit magnitude and exposure time of each band. The limit magnitude is based on the values given per pixel by {\tt SExtractor} and are calculated for a $2\arcsec$ (diameter) circular aperture at $5 \sigma$. The minimum number of images stacked for each filter per pointing is four.

For approximately 7\,deg$^2$ of the XMM LSS area we made use of existing data reduced by the CFHTLenS collaboration \citep{Heymans2012} using similar tools and methods to our approach, which ensure uniformity in the final data products \citep{Hildebrandt2012,Erben2013}. 

\section{Data reduction \& source selection} \label{data_reduction}
\subsection{Basic data reduction} 

The CFHT data retrieved from the archive are already pre-processed with {\tt ELIXIR}\footnote{CFHT data reduction pipeline} \citep{Magnier2004}. This pre-processing includes the masking of dead or hot pixels, bias and overscan correction, flat-fielding, photometric superflat, fringe correction for the $i$ and $z$ data, and a rough astrometric and photometric solution for each field. 

We detail below the main steps of the subsequent data reduction process, which are based on the work-flow used by the CFHTLenS collaboration \citep{Heymans2012,Erben2013}, additionally convolving the different bands to the same (worst) seeing (PSF homogenization) \citep{Hildebrandt2013}.

\begin{enumerate}
\setlength\itemsep{0.5em}
 
  \item A basic quality control was carried out for each of the images, identifying chips with a large number of saturated pixels, severe tracking errors, misidentified image type, incorrect exposure time etc.
  
  \item Satellite tracks were identified using a method based on a feature extraction technique \citep[Hough transform, see][]{Vandame2001}.
  
  \item Weight images were created for each chip, including dead or hot pixels or columns, saturated areas of the chips (e.g. the centres of very bright stars) and the satellite track masks from the previous step.
  
  \item The source catalogues necessary for astrometric calibration were created using {\tt Source Extractor} \citep[{{\tt SExtractor},}][]{Bertin1996}.
  
  \item Absolute, internal astrometric calibration, and the relative photometric calibration of the $ugriz$-band images was accomplished for each field using {\tt Software for Calibrating AstroMetry and Photometry (SCAMP)} \citep{Bertin2006} and the 2MASS astrometric catalogue \citep{Skrutskie2006} as a reference. 
  
  \item The coaddition of images was accomplished using the weighted average method with the {\tt SWarp} software \citep{Bertin2002}.
  
  \item To account for the photometric issues created by PSF heterogeneity between different bands, we convolved the images to the same PSF using methods developed by \cite{Kuijken2008}. 
  
  \item With {\tt  SExtractor} it is possible to detect sources in one band and measure photometric quantities on another (dual image mode). We detected sources on the $r$-band, which is on average the deepest. This has the advantage that photometry can be forced in another band at a location where a source is known to exist and that colours are very accurately estimated if the PSF is uniform between different bands. The multicolour catalogue contains measurements in all bands for all of the $r$-band detected sources, in isophotal apertures defined by the $r$-band measurement. Five contiguous pixels with a detection threshold of $1.5\,\sigma$ above the background are the minimum criteria required to have a detection by {\tt SExtractor}.

  \item To mask image defects and regions where photometry is unreliable (around bright stars because of halos and diffraction spikes, in areas with a low signal-to-noise ratio, around reflections producing ghost images of bright objects, on top of asteroid tracks etc.), we used the {\tt AUTOMASK} software \citep[see][]{Erben2009,Dietrich2007} and information from the image weights for all bands used in selecting the $u$-dropouts. Furthermore, each image was individually inspected visually and other problematic regions were manually excluded from the analysis. The masked objects were flagged in the multi-colour catalogue.
   
  \item The final absolute photometric calibration was based on SDSS DR10 \citep{SDSSDR10}. We compared the median magnitude of stellar objects in our multicolour catalogue and shift each band to match with the median magnitude of the same objects in SDSS DR10.
  
  \item Photometric redshifts were estimated using the {\tt BPZ}\footnote{\url{http://acs.pha.jhu.edu/~txitxo/bpzdoc.html}} code \citep{Benitez2000}, based on priors from the VIMOS VLT Deep Survey \citep[VVDS, see][]{VVDS,Raichoor2014}. We also provide photo-z estimates for objects that are not detected in one or more of the $ugiz$-bands (objects that have magnitudes fainter than the limit magnitudes in each field, which can occur with the dual-image mode of {\tt SExtractor}). We note though that photometric redshifts were not used in this study.
\end{enumerate}

The co-added images, weights, masks, associated source catalogues and systematic effects check-plots can be provided on request from the authors. 

\begin{table}[ht]\setlength\extrarowheight{2.5pt} 
  \centering
    \caption{Average seeing (before PSF homogenization), limit magnitude and exposure times for each filter of the CFHT individual pointings.}
   \label{Table:data_properties}
  \begin{tabular}{cccc}
  \hline
  \hline
  \hspace{2mm}
    Filter & Seeing &  Limit magnitude &  Exposure time \\
           & (\arcsec)    & (mag)            & (hours) \\
    \hline       
    u & $ 0.96  $ & $ 24.28  $ & $ 1.17 $ \\
    g & $ 0.95  $ & $ 24.61  $ & $ 0.91 $ \\
    r & $ 0.81  $ & $ 24.20  $ & $ 0.87 $ \\
    i & $ 0.80  $ & $ 23.50  $ & $ 0.59 $ \\
    z & $ 0.68  $ & $ 23.15  $ & $ 1.76 $ \\
    \hline
  \end{tabular}
\end{table}
   
\subsection{LBG candidates}
\label{Section:LBG}   
The background population of sources used to probe the magnification signal consists of $u$-dropouts which are LBG candidates. LBGs are high-redshift  galaxies that undergo star formation at a high rate \citep{Steidel1998}. Because radiation at higher energies than the Lyman limit is almost completely absorbed by the neutral gas surrounding star-forming regions, their apparent magnitude changes abruptly for a combination of filters spanning the Lyman limit. Employing a combination of filters in the optical domain, generally one can select LBGs at a redshift $z>2.5$.

LBGs have been used successfully in the past for magnification studies \citep[see][]{Hildebrandt2009, Hildebrandt2013, Morrison2012, Ford2012,Ford2014}  since their luminosity function and redshift distribution are relatively well understood. Because the magnification signal is sensitive to the slope of the number counts of the sources used, knowledge of the luminosity function is essential for such measurements. Another advantage of using LBG as background sources is that they are situated at much higher redshifts than the galaxy clusters studied here, therefore reducing the probability of having a magnification-like signal induced by physical correlations between sources and clusters.

\begin{figure}
 \centering
 \includegraphics[width=92mm]{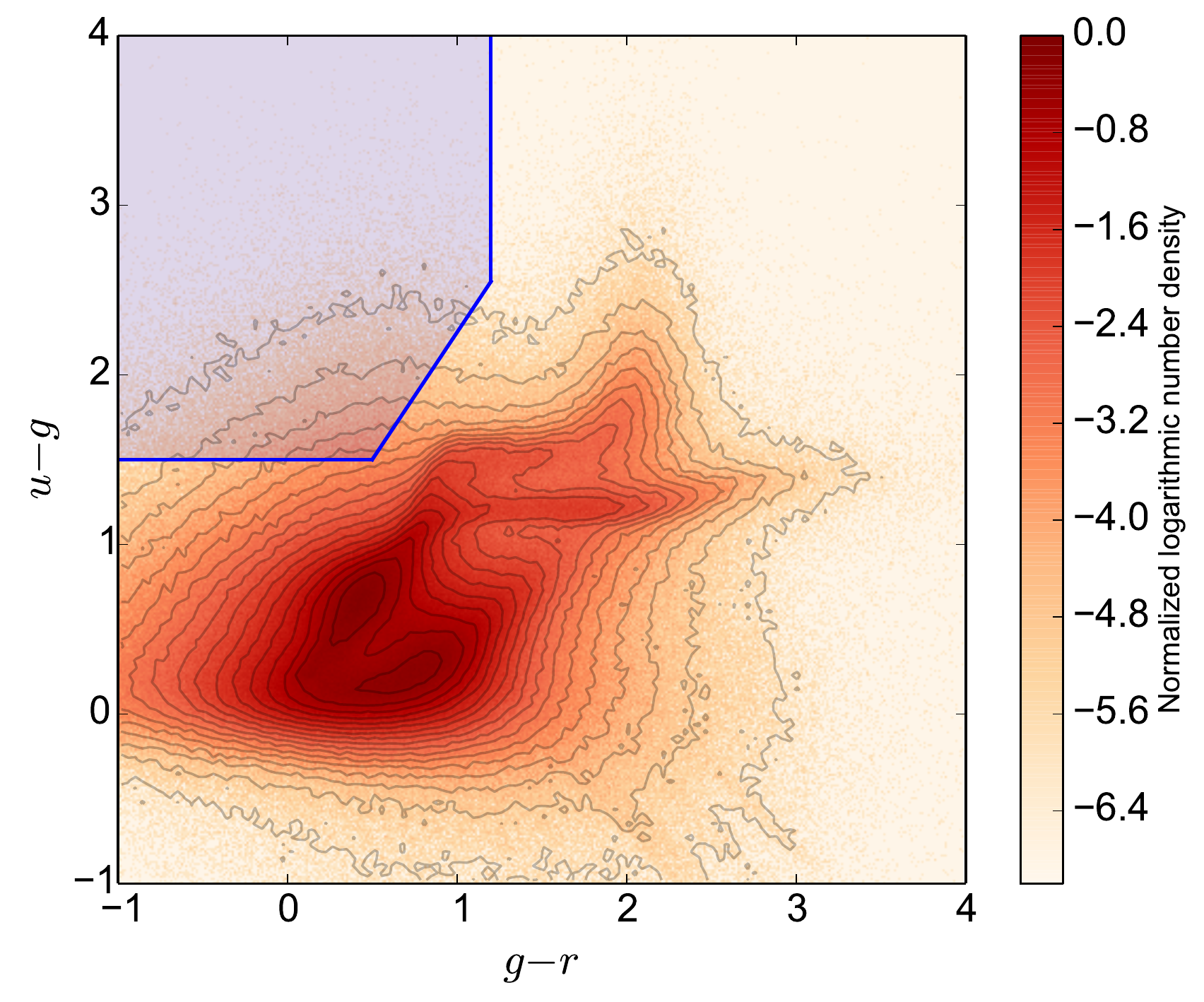}
 \caption{$u-g$ \,vs\, $g-r$ colour-colour number density plot of the galaxies in the SpARCS fields, selected with the {\tt SExtractor} parameter CLASS\_STAR $< 0.9$. The colour selection criteria described in Sect. \ref{Section:LBG} are delineated on the upper left of the image with the shaded area and the blue lines.}
 \label{fig:grug}
\end{figure}

 For the $u$-dropouts, we adopted the colour selection criteria previously used in \cite{Hildebrandt2009}: 
\begin{equation*}  
\left\{ \begin{array}{ll}
1.5 <(u-g) &  \\ 
-1.0 < (g-r)<1.2 & \\ 
1.5 \, (g-r) < (u-g)-0.75 . & \\ 
\end{array} 
\right. 
\end{equation*}

\begin{figure}
 \centering
\includegraphics[width=75mm]{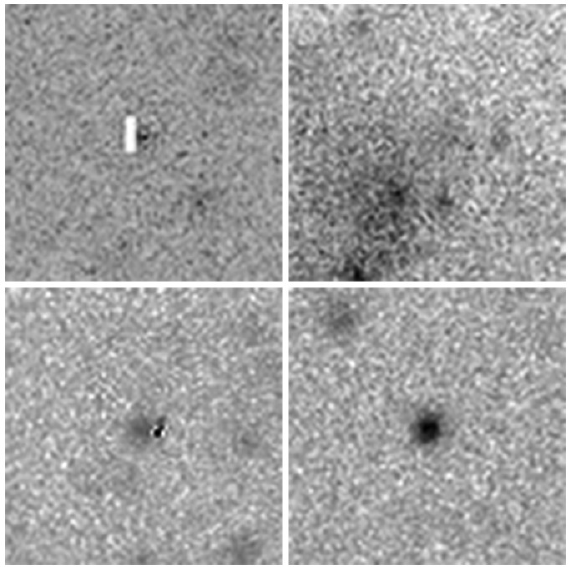}
   \caption{Examples of LBG candidates rejected after the visual inspection of the entire sample (top row and bottom left) and one example of an accepted u-dropout (bottom right). Top left and bottom left candidates were rejected due to hot and cold pixels respectively, while the top-right candidate was rejected because of the diffuse light contaminating the photometry.}
 \label{fig:udrop_cleaning}
\end{figure}

\begin{figure}
 \centering
\includegraphics[width=91mm]{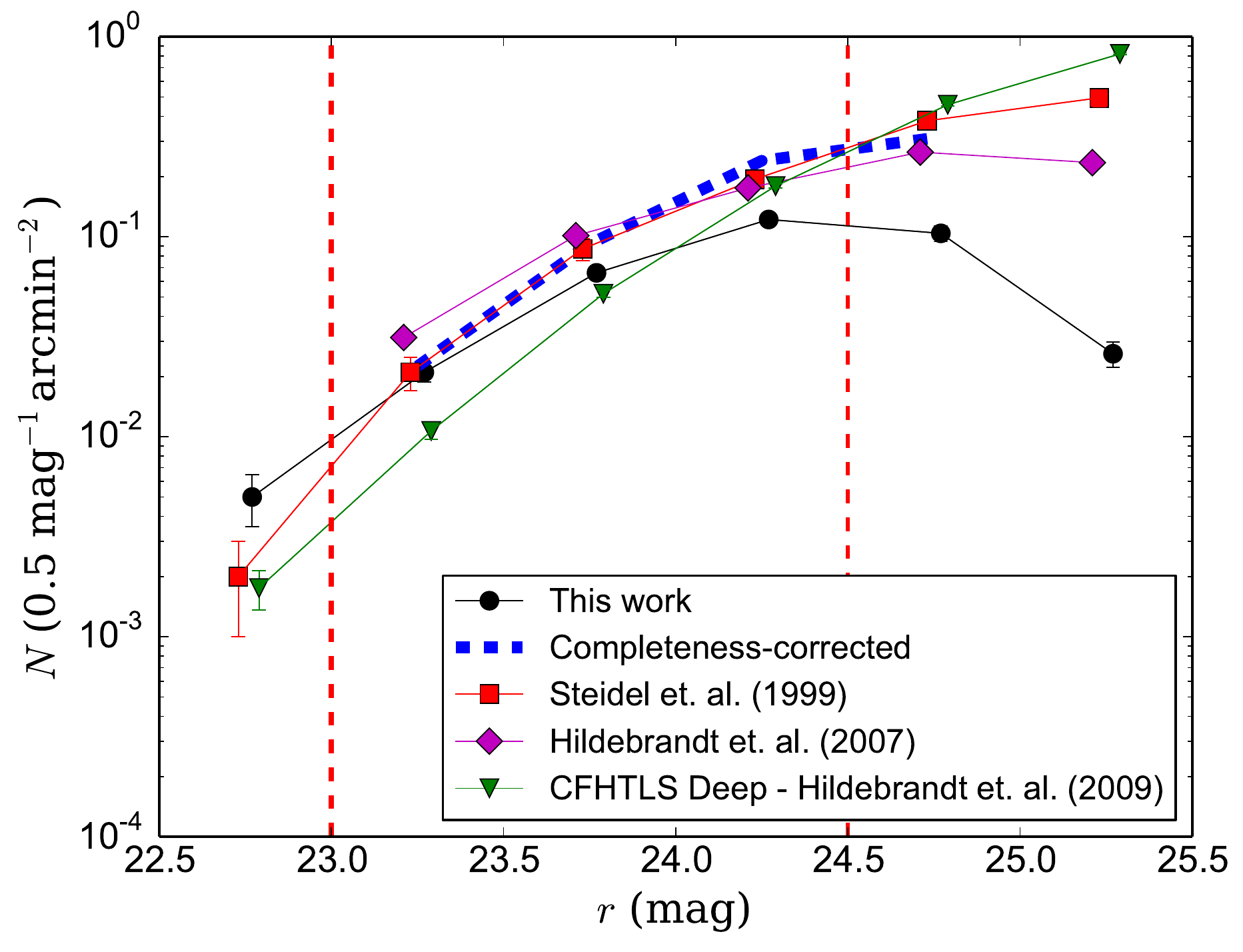}
   \caption{Number counts of the SpARCS $u$-dropout sample compared to previous work at wavelengths that roughly match the same rest-frame in the UV. The blue-dashed line represents the completeness-corrected $u$-dropouts number counts we measure.}
 \label{fig:udrop}
\end{figure}

\begin{figure}[ht]
 \centering
\includegraphics[width=90mm]{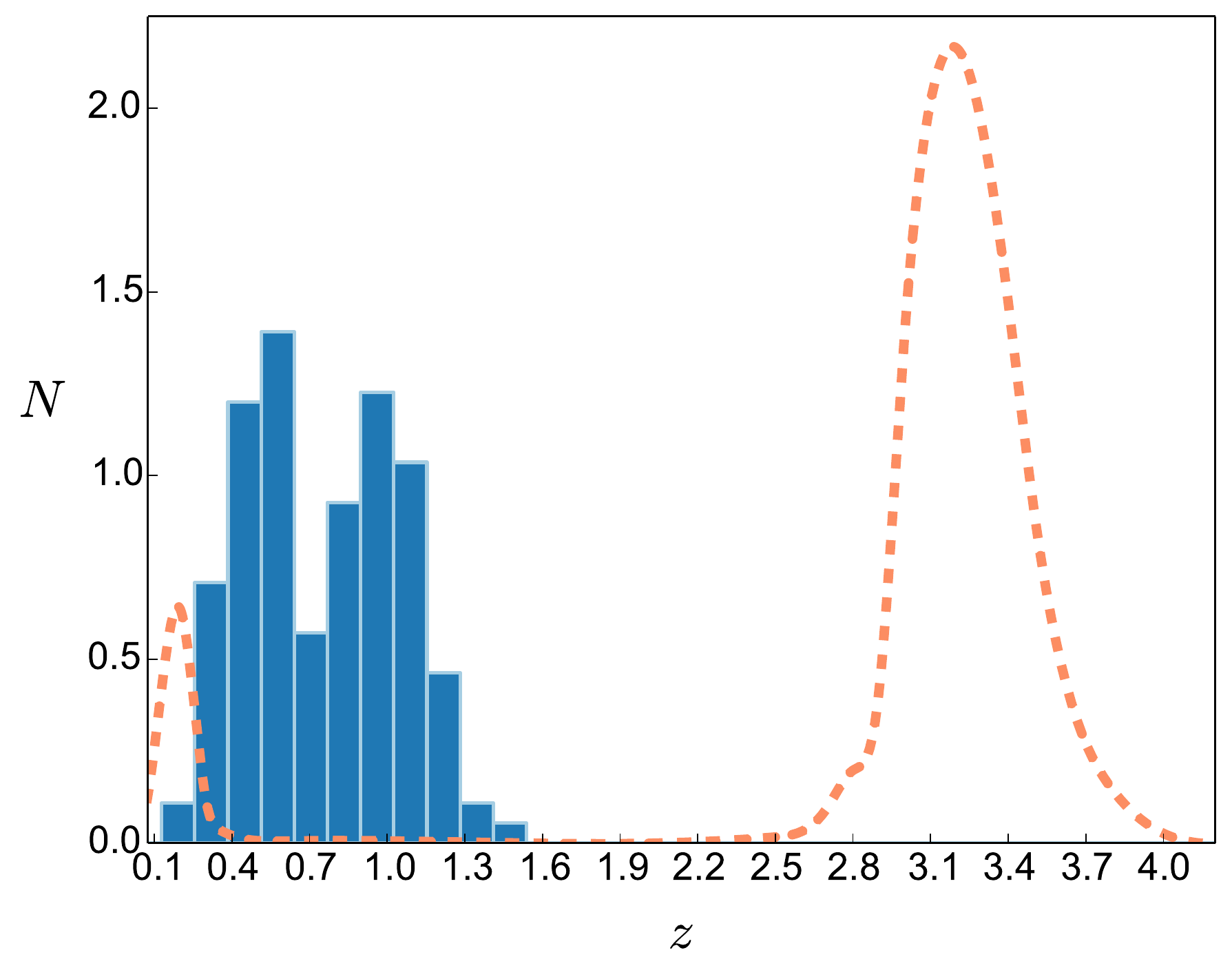}
 \caption{Histogram of the redshift distribution of SpARCS clusters (with total counts normalized to unity) and the redshift probability distribution function of the LBG candidates (orange dashed line).}
 \label{fig:cluster_LBG_z_hist}
\end{figure}

Figure \ref{fig:grug} shows the distribution of the number density of galaxies in the $u-g$ vs $g-r$ colour space, with contours logarithmically spaced. The selected $u$-dropouts are located in the shaded area. The selection of dropouts using these cuts in the $g-r,\, u-g$ colour space has been shown with simulated data (for a similar, but deeper data set) to produce a contamination level from stars and low-z interlopers below 10\% for each magnitude bin \citep{Hildebrandt2009}. We also required the candidates to have a {\tt SExtractor} CLASS\_STAR parameter smaller than 0.9, which facilitates the rejection of most stars in the sample. Since our median FWHM is $0\farcs8$ in the detection band, we could still reliably separate stars from high-redshift galaxies at bright magnitudes. An additional size constrains was added, requiring the object to be smaller than 5$\arcsec$, since LBGs at $z=3.1$ have a maximum size of about $2-3$$\arcsec$ \citep{Giavalisco1996b}. Furthermore, after applying the image masks to the data, each object in the resulting sample of LBG candidates was visually inspected, rejecting obvious false detections such as:
\begin{itemize}
\item very extended objects
\item bright knots in spiral galaxies
\item densely populated fields (numerous objects, partially or completely overlapping)
\item other image defects not being masked automatically
 \end{itemize}
 A few examples of rejected $u$-dropout candidates can be seen in Fig. \ref{fig:udrop_cleaning}.
 
 A comparison between the $u$-dropout number counts as a function of magnitude in our sample with other work can be seen in Fig. \ref{fig:udrop}. We estimated the fraction of LBGs that are lost due to the limited depth of the data from simulations similar to the ones used in \citep{Hildebrandt2009}. We created mock catalogues of SpARCS depth as well as CFHTLS-Deep depth, the latter of which are highly complete down to the magnitude limit of SpARCS. Using the ratio of the number counts between the two catalogues as an incompleteness correction for fainter magnitude bins, the number density of dropouts matches very well with other measurements in the literature. We note that this correction was just used for this figure and not used in the subsequent analysis. Due to the large survey volume, the cosmic variance contribution can safely be neglected.
 
Applying the magnitude cuts and masks to the catalogues, and after the manual rejection of false LBGs, we selected 16\,242 $u$-dropouts with magnitudes in the interval range 23 $\leq r \leq$ 24.5, located at a mean redshift of $z$ $\sim$ $3.1$. This magnitude range was chosen to minimize as much as possible low-redshift contamination, while still having a sufficient number of galaxies for a meaningful measurement. 

Another peculiarity of using LBG as sources is that we had to model the redshifts of contaminants to be able to minimize their influence on the measurements. As seen in Fig. \ref{fig:cluster_LBG_z_hist}, there is practically no overlap between sources and lenses at high redshifts. Additionally, we measure the cross-correlation for a sample of clusters that does not include the low-redshift region $ z < 0.3$, to control for, and reduce the possibility of having a positive signal from low-redshift, physically-induced cross-correlations. We found that since there are very few clusters with $0.2 \leq z \leq 0.3$, there is almost no difference if we either include or exclude them from the measurements. 

Detailed properties of the LBG populations selected using the same methods have been described by \cite{Ford2014} and \cite{Hildebrandt2013}. 

\section{Methods} \label{methods}
\subsection{Masking correction}
\label{Section:masking_correction}
Another effect that could disrupt our measurement would be the fact that galaxy cluster candidate galaxies are effectively masking some of the LBG candidates in the background. \cite{Umetsu2011a} have developed a method of estimating the amount of masking based on deep Subaru imaging data for a sample of 5 massive clusters ($\geq  10^{15}$\,M$_{\sun}$) at intermediate redshifts ($0.18 \leq  z \leq  0.45$). The study found that while at large radii the masking is insignificant, amounting to only a few percent, at small radii the cluster galaxies can occupy even 10-20 percent of the annulus area. 

\begin{figure}
 \centering
\includegraphics[width=91mm]{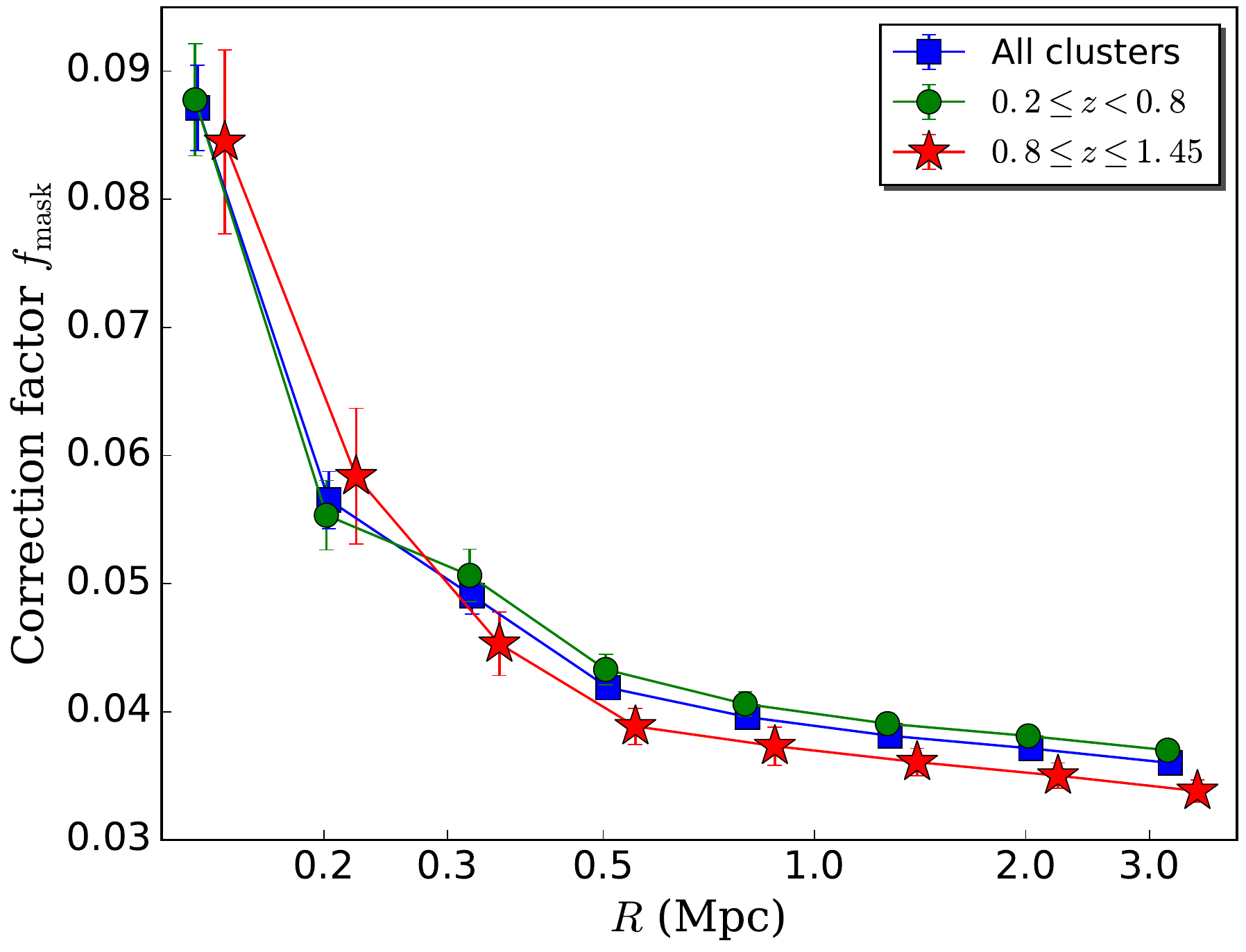}
 \caption{Masking correction factors $f_{\mathrm{mask}}$ as a function of the  redshift of cluster samples. The data points are slightly shifted on the x-axis for the sake of clarity.}
 \label{fig:masking_correction_z}
\end{figure}

To correct for this additional masking, we adopted a simple method similar to that described by \cite{Umetsu2011a} in their Appendix A. We selected all objects brighter than $r$-band magnitude 24.5 (the LBG candidates' magnitude limit) and fainter than $r$-band magnitude 16 (where our automatic masking procedure would have already masked the objects). The area of every object was taken to be the isophotal area above the detection threshold of $2.5 \sigma$. For each cluster, the area of the objects was summed at every corresponding radial bin to calculate the proportion of area covered, with which the magnification signal was boosted. For all cluster samples, we  average the correction factors $f_{\mathrm{mask}}$ and take the errorbars as their $1 \sigma$ standard deviation. 
Figure \ref{fig:masking_correction_z} shows that the masking fraction depends only slightly on the redshift of lenses, while for clusters of different richness we do not find a significant variation of the masking fraction amplitude. 

Although \cite{Umetsu2011a} find almost twice the amount of masking we find at small annuli, this difference can most likely be explained by the slight differences in methodology, by the fact that the only cluster for which they have published the masking correction is a highly unusual one (the very massive and rich Abel 1689) and because at low redshift the galaxies are larger down to a given surface brightness. 

\subsection{Magnification of number counts} \label{magnification_theory}

In terms of $\kappa$ and $\gamma$, the convergence and shear, we can describe to first order the image deformation from the source to the observer frame through the Jacobian matrix $\cal{A}$:
\begin{equation}
\cal{A} = \left( \begin{array}{cc}
{1-\kappa-\gamma_1} & {-\gamma_2} \\
{-\gamma_2} & {1-\kappa+\gamma_1} \\
\end{array} \right) \,.
\end{equation}
The magnification factor $\mu$ is the inverse of the determinant \citep{Bartelmann2001}:
\begin{equation}
\mu = \frac{1}{\mathrm{det} \cal{A}} = \frac{1}{(1-\kappa)^2 - \left|\gamma\right|^2} \,,
\end{equation}
where $\left|\gamma\right|^2 = \sqrt[]{\gamma_1^2+\gamma_2^2}$; $\gamma_1$ and $\gamma_2$ representing the shear components.
 
The magnification produced by a gravitational lens can be detected through the change from inherent ($N_0$) to observed ($N$) differential number counts of background sources: 
\begin{equation}
N(m) \, \mathrm{d}\,m = \mu^{\alpha-1} N_0(m) \, \mathrm{d}m \,,
\end{equation}
 \citep{Narayan1989}, where $m$ is the apparent magnitude of sources, and $\alpha \equiv \alpha(m)$ {is proportional to the logarithmic slope of the source number counts as: 
 \begin{equation}
\alpha = \alpha(m) = 2.5 \frac{\mathrm{d}}{\mathrm{d}m}\log N_0(m) \,.
\end{equation}
This means that the observed spatial source density} of lensed galaxies can either increase or decrease, depending on the sign of $\alpha - 1$. Galaxies with number counts where $(\alpha - 1) > 0$ will appear to be spatially correlated with the position of lenses, while for $(\alpha - 1) < 0$ an anti-correlation will be observed. There is no effect in the case of $(\alpha - 1) \approx 0$, since the dilution and amplification effect will mutually cancel out.
 
    For the cross-correlation measurement between the positions of galaxy clusters and LBG candidates, we assigned a weight factor for each source of $\alpha-1$, according to its position on the luminosity function (magnitude) \citep{Scranton2005,Menard2003}. 
 
    To estimate the optimal weight factor $\alpha - 1$ required for both the measurement and its interpretation, we relied on external LBG luminosity function measurements for the characteristic magnitude $M^\ast$ and faint-end slope $\alpha_{\mathrm{LF}}$ \citep{vanderBurg2010}. For the $u$-dropouts $M^\ast =  -20.84$ and $\alpha_{\mathrm{LF}} = -1.6$
\begin{equation}
\alpha = 10^{0.4(M^\ast-M)}-\alpha_{\mathrm{LF}}-1 \,.
\end{equation}
LBGs selected using our method occupy a narrow region in redshift space centred around $z \approx 3.1$, which we approximate with a Dirac Delta function at the centre of the distribution. The validity of the approximation is supported by the fact that the angular diameter distance, on which the lensing signal depends, does not change significantly over the narrow range of the distribution. 
    
\subsection{Magnification model}
   
   The magnification signal from galaxy clusters was modelled using the Navarro-Frenk-White (NFW) profile and a two-halo term from large-scale structure, as well as taking the effects of halo miscentering into account. 

The convergence and shear were modelled as the sum of two terms:
 \begin{equation}
 \kappa(z) = \left[\Sigma_{\mathrm{NFW}}(z) + \Sigma_{\mathrm{2halo}}(z) \right]/\Sigma_{\mathrm{crit}} ,
 \label{modelEQkappa}
 \end{equation}
 \begin{equation}
 \gamma(z) = \left[\Delta\Sigma_{\mathrm{NFW}}(z) + \Delta\Sigma_{\mathrm{2halo}}(z) \right]/\Sigma_{\mathrm{crit}} \,,
 \label{modelEQgamma}
 \end{equation}
where $\Sigma_{\mathrm{crit}}(z)$ is the critical surface mass density at the lens redshift, $\Sigma_{\mathrm{NFW}}$ is the surface mass density from the NFW halo, $\Sigma_{\mathrm{2halo}}$ corresponds to the critical surface mass density from the two-halo term and $\Delta\Sigma$ represents the differential surface mass density.
 The full expressions for the the surface mass density and differential surface mass density dependence on the dimensionless radial distance $x = R/r_{s}$ of an NFW lens are given by \citet{Bartelmann1996,Wright2000}.

The critical surface mass density can be described in terms of the angular diameter distances between observer-lens $D_\mathrm{l}$, observer-source $D_\mathrm{s}$ and between lens-source $D_\mathrm{ls}$, the gravitational constant $G$ and the speed of light $C$ (not to be confused with the concentration parameter, $c$):
\begin{equation}
\Sigma_{\mathrm{crit}} = \frac{C^2}{4\pi G} \frac{D_\mathrm{s}}{D_\mathrm{l} D_{\mathrm{ls}}} \,.
\end{equation}

The NFW density profile is given by:
 \begin{equation}
\rho(r) = \frac{\delta_{\mathrm{c}} \, \rho_{\mathrm{crit}}(z)}{\left(r/r_{\mathrm{s}}\right)
 \left(1+r/r_{\mathrm{s}}\right)^{2}} \,,
 \label{rhonfw}
 \end{equation}
 where $\rho_{\mathrm{crit}}(z)$ is the critical density at the halo redshift $z$:
 \begin{equation}
\rho_{\mathrm{crit}}(z) = \frac{3H^{2}(z)}{8\pi G} \,.
 \label{cnfw}
 \end{equation}
 $H(z)$ is the Hubble parameter at the same redshift,  $G$ is Newton's constant, the scale radius is given by $r_{\mathrm{s}} = r_{\mathrm{200}}/c$, where $c$ is the dimensionless concentration parameter, and the characteristic halo over-density 
 $\delta_{\mathrm{c}}$ is given by:
 \begin{equation}
 \label{eq:characteristic_halo_overdensity}
\delta_{\mathrm{c}}= \frac{200}{3}\frac{c^{3}}{\ln(1+c)-c/(1+c)} \,.
 \end{equation}
 The radius $r_{\mathrm{200}}$ is defined as the radius inside which the mass of the halo is equal to 200 $\rho_\mathrm{crit}$ \citep[see][]{Navarro1997}. 

$\Sigma_{\mathrm{2halo}}$ quantifies the contribution of neighbouring halos to the surface mass density and is given by \cite{Johnston2007} as:
\begin{equation}
 \Sigma_{\mathrm{2halo}}(R,z) = b_{\mathrm{l}} \, (M_{\mathrm{200}},z) \,  \Omega_\mathrm{M}  \, \sigma_8^2 \,  D(z)^2  \, \Sigma_\mathrm{l}(R,z)\,,
\end{equation}
with
\begin{equation}
\Sigma_\mathrm{l}(R,z) = (1+z)^3 \rho_{\mathrm{crit}}(0) \int_{-\infty}^\infty \xi\left( (1+z)\sqrt{R^2 + y^2} \right) \mathrm{d}y\,,
\end{equation}
and
\begin{equation}
\xi(r) = \frac{1}{2\pi^2} \int_0^\infty k^2 P(k) \frac{\sin{kr}}{kr} \mathrm{d}k \,,
\end{equation}
where $r$ is the comoving distance, $D(z)$ is the growth factor, $P(k)$ is the linear matter power spectrum, and $\sigma_8$ is the amplitude of the power spectrum on scales of 8 $h^{-1}$Mpc. The lens bias factor $b_{\mathrm{l}}$ is approximated by \cite{Seljak2004} with:
\begin{equation}
b_\mathrm{l} (x=M/M_{\mathrm{nl}}(z)) = 0.53 + 0.39 x^{0.45} + \frac{0.13}{40x+1} +5 \cdot 10^{-4} x^{1.5} \,,
\end{equation}
where the nonlinear mass $M_{\mathrm{nl}}$, is defined as the mass within a sphere for which the root mean square fluctuation amplitude of the linear field is 1.68.

Cluster miscentering was taken into account statistically in the model by shifting the cluster centers with a radial offset following a 2-D Gaussian probability distribution \citep[see Fig. 1 in][]{Ford2014}. This had the net effect of smoothing the surface mass density at small scales for the NFW-2halo term model used.

The cross-correlation $w(R)$ between the position of galaxy cluster centres and positions of LBG candidates was measured in seven logarithmic physical radial bins to 3.5\,Mpc. 

By stacking in physical radial bins instead of angular bins, we ensured that mixing clusters of different redshifts does not stack the magnification signal from different physical scales. We measured the magnification signal for each cluster sub-sample, each time drawing randoms 1000 times the size of the sources catalogue and with the same masking layout to account for the survey geometry. Since we only had one single measurement per pointing, we did not draw random catalogues for the clusters as well, summing instead the pairs for each angular bin for all clusters in the sample:
\begin{equation}
\label{wopt}
\mathrm{w}(R) = \frac{S^{\alpha -1}L-\langle\alpha -1\rangle LR_*}{LR_*} \,,
\end{equation}
where $L$  stands for lenses, $S^{\alpha -1}$ for optimally-weighted sources and $R_*$ for randoms. The terms represent normalized pair counts in physical radial bins. 

 Full covariance matrices are estimated for each set of independent measurements directly from the data (see Fig. \ref{fig:cov_matrix} for the covariance matrix of the entire sample of measurements). 

Assigning a constant weight for all LBG changes results only very slightly since the slope of the number counts does not change much over the magnitude interval where we perform our measurements.

To avoid entering the strong lensing regime in the innermost regions of clusters, we restricted our measurements and the model to radii larger than 1.5 times the Einstein radius. For convenience, we use the Einstein radius $\theta_{\mathrm{E}}$ for an isothermal sphere, which is given by:

\begin{equation}
\label{Einstein_radius_isothermal}
\theta_{\mathrm{E}} = 4\pi \left(\frac{\sigma_{\mathrm{v}}}{C}\right)^2\frac{D_{\mathrm{ls}}}{D_{\mathrm{s}}},
\end{equation}

where $\sigma_{\mathrm{v}}$ is the velocity dispersion in $\mathrm{km \, s^{-1}}$, calculated using Eq. 1 of  \cite{Munari2013}:

\begin{equation}
\label{velocity_dispersion}
\sigma_{\mathrm{v}} = 1100 \cdot \left[\frac{h(z) M_{200}}{10^{15} \mathrm{M_\sun}} \right]^{1/3},
\end{equation}
where h(z) is the dimensionless Hubble parameter at redshift z. 

We calculated $\theta_{\mathrm{E}}$ for each cluster and discarded the measurements performed at radii smaller than 1.5 times of this value. As $\theta_{\mathrm{E}}$ is usually smaller than the innermost bin edge, only a small proportion of the measurements is lost this way. We accounted for this by restricting the model to the same radii as the data. This is necessary because the mass-richness relation we use for calibration results in clusters massive enough to have their $\theta_{\mathrm{E}}$ within our measurements range, which induces model instability and artefacts.  

\subsection{Signal-to-noise ratio estimates}\label{Section:signal_to_noise}
To estimate the expected signal-to-noise ratio we used the methods derived by \cite{vanWaerbeke2010b}. As the signal-to-noise ratio is so low for most individual clusters that direct measurements of the signal are impossible, we relied on stacking multiple foreground lenses to decrease the noise of the average magnification as a function of the distance from cluster centres. The average mass and concentration parameters ($M_{200}$ and $c_{200}$) of the lenses  that contribute to the average magnification profile can then be constrained with the likelihood:
\begin{equation}
{\cal L}\propto \exp
\left[(\delta_\mathrm{N}(\theta)-\bar\delta_\mathrm{N}(\theta))C^{-1}_{{\delta_\mathrm{N}}{\delta_\mathrm{N}}}(\delta_\mathrm{N}(\theta)-
\bar\delta_\mathrm{N}(\theta))^{\rm\bf
T}\right] \,,
\end{equation}
 where $\delta_\mathrm{N}(\theta)$ is the mean galaxy radial counts contrast profile that we are measuring, and $\bar\delta_\mathrm{N}(\theta)$ is the galaxy count profile model. The noise covariance matrix, $C_{{\delta_\mathrm{N}}{\delta_\mathrm{N}}}$, was estimated by choosing 200 random positions for which we estimated their angular correlation function with our LBG candidates sample. As expected, the angular correlation is consistent with zero and the dispersion around the mean corresponds to the $C_{{\delta_\mathrm{N}}{\delta_\mathrm{N}}}$ matrix elements. We then scaled the amplitude of the noise covariance matrix to match the actual number of lenses that we use. This method ensures that the halo sampling and clustering noise of the source population are appropriately taken into account.
 By maximizing the likelihood function we found an expected signal-to-noise ratio of about 10 for using all lenses in our sample and $6$ for halos with $z \geq 0.80$.
 
\subsection{Composite-halo fits}
\label{multihalo}

\begin{figure}
 \centering
\includegraphics[width=92mm]{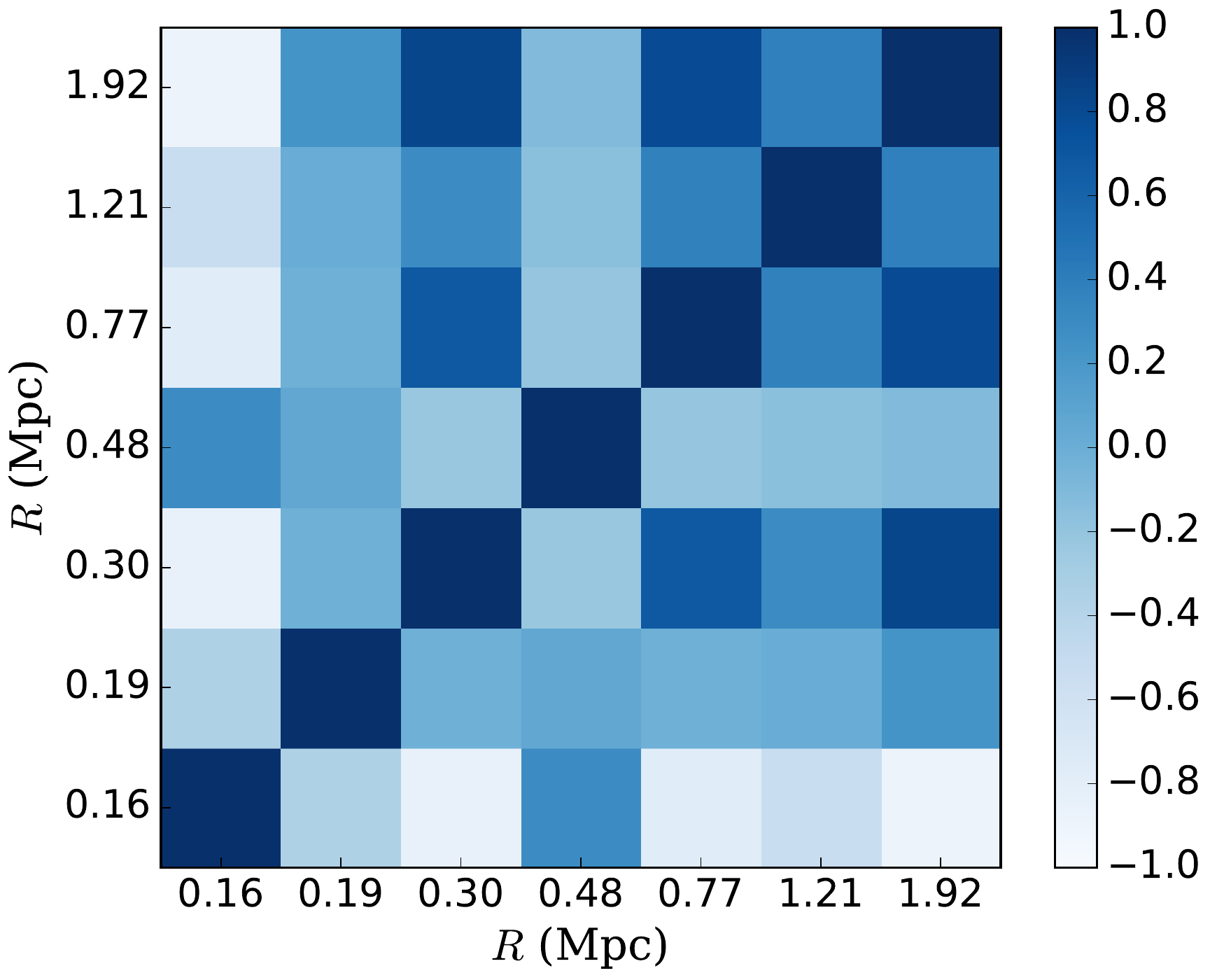}
 \caption{Correlation matrix (normalized covariance matrix) of the optimally-weighted cross-correlation function between $u$-dropouts and the centres of galaxy clusters. }
 \label{fig:cov_matrix}
\end{figure}

The weak lensing magnification contribution to the cross-correlation signal can be calculated as follows:
\begin{equation}
\label{wmodel}
\mathrm{w}_{\mathrm{lensing}}(R) = \frac{1}{N_{\mathrm{lens}}} \sum_{i=1}^{N_{\mathrm{lens}}} \, \langle(\alpha-1)^2\rangle_i \, \left[\mu \, (R,M_{200})_i-1\right] \,,
\end{equation}
where the sum is over the number of clusters (lenses) in a given sample selected for stacking.
 
 Our results are not too sensitive to the choice of the mass-concentration relation and we chose the one developed by \cite{Duffy2008} to fix the halo concentration parameter $c$ (see Equation \ref{eq:characteristic_halo_overdensity}):
 \begin{equation}
c = A_{200}(M_{200}/M_{\mathrm{pivot}})^{B_{200}}(1+z_{\mathrm{cluster}})^{C_{200}} \,,
\label{eq:prada_mass_concentration_relation}
\end{equation}
where  $A_{200} = 5.71$,  $B_{200} = -0.084$,  $C_{200} = -0.47$,  $M_{\mathrm{pivot}} = 2\times 10^{12}$ and $z_{\mathrm{cluster}}$ is the redshift of the lens.

 However, since we needed a mass to fix the halo concentration parameter, we first made use of the mass-richness relation determined from the GCLASS cluster sample:
\begin{equation}
M_{200} = a \, N_{\mathrm{red}}^{2.97\pm\,0.26} \times 10^{10.63\pm\,0.39} \,.
\label{eq:GCLASS_mass_richness_relation}
\end{equation}
The richness values for the GCLASS sample are Eddington-biased high, given that the GCLASS sample was selected from the main SpARCS sample due to high $N_\mathrm{red}$ values.

The sole fit parameter in our measurement was the amplitude of the scaling relation between mass and richness $a$ (see Equation \ref{eq:GCLASS_mass_richness_relation}). We used only the central value of Eq. \ref{eq:GCLASS_mass_richness_relation} for the fitting procedure. The fit was performed by varying $a$ and minimizing the reduced $\chi^2$ between the magnification model described above and the cross-correlation measurement.  
 We also added for comparison in Fig. \ref{fig:mass_richness_relation} the mass-richness relation determined by \cite{Muzzin2007b} for the Canadian Network for Observational Cosmology \citep[CNOC1, see][]{Yee1996} set of clusters, a survey of 16 rich galaxy clusters with $0.17 < z < 0.55$:
\begin{equation}
M_{200} = (69.4 \, N_{\mathrm{red}})^{1.62\pm\,0.24} \times 10^{9.86\pm\,0.77} \,.
\label{eq:muzzin_mass_richness_relation}
\end{equation}
We used the full covariance matrix (shown in Fig. \ref{fig:cov_matrix}), as determined from the measurements themselves, to find the minimum reduced $\chi^2$. 

Assuming statistically-independent data points (bins) and idealized Gaussian noise, the inverse of the maximum-likelihood estimator of the covariance is biased, with an amount depending on the ratio between the number of bins and independent measurements.
\citet{Hartlap2007} have determined a correction factor (see their Eq. 17) which we applied here to avoid underestimating the error bars.

\section{Results} \label{results}

\begin{figure*}
 \centering
\includegraphics[width=138mm]{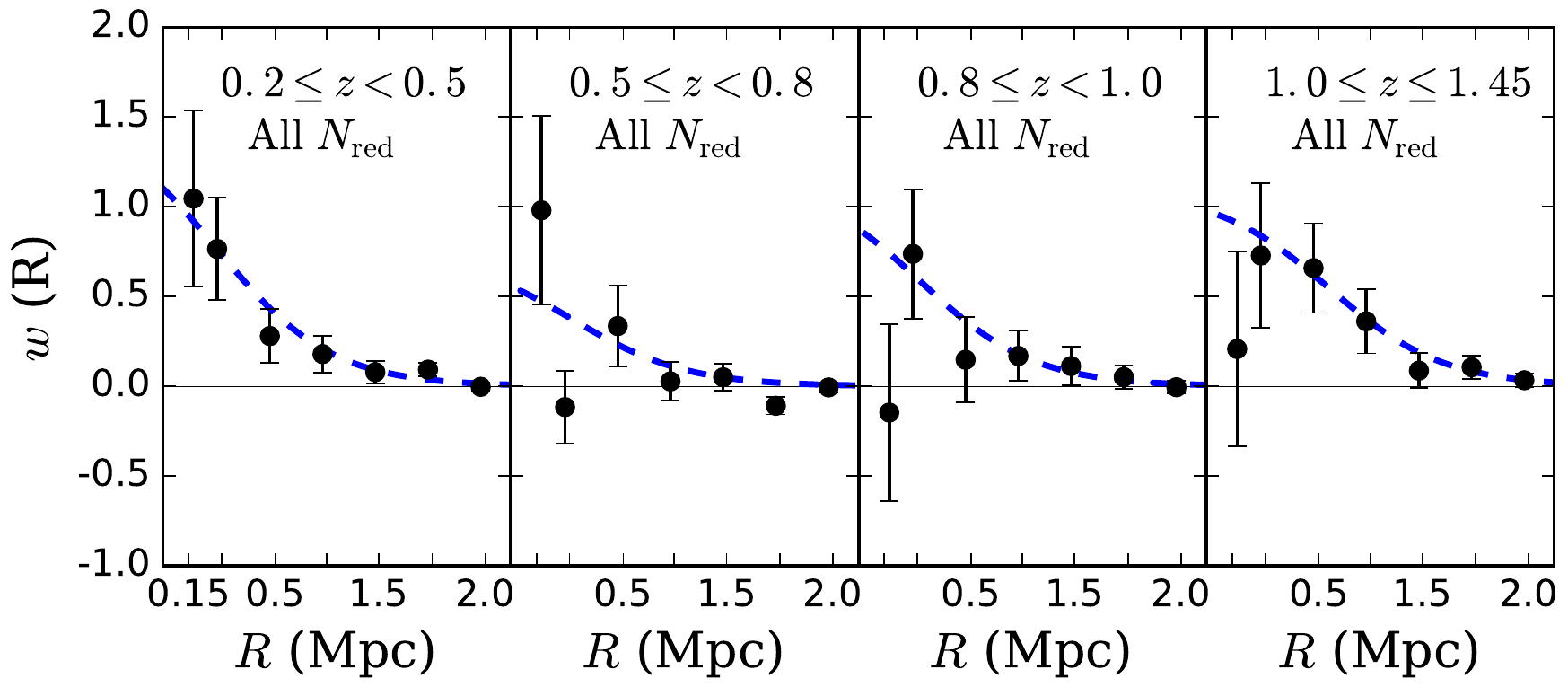}
 \caption{Angular cross-correlation measurements between the $u$-dropouts and the centres of galaxy clusters, as a function of the radial distance from the cluster centres in physical units. The sample of galaxy clusters on which the measurement was performed is binned in redshift as shown in each of the figure legends. Best-fit models are plotted with the dotted blue line, while our measurements are represented by the black round points.}
  \label{fig:correlation_z}
\end{figure*}

\begin{figure*}
 \centering
\includegraphics[width=138mm]{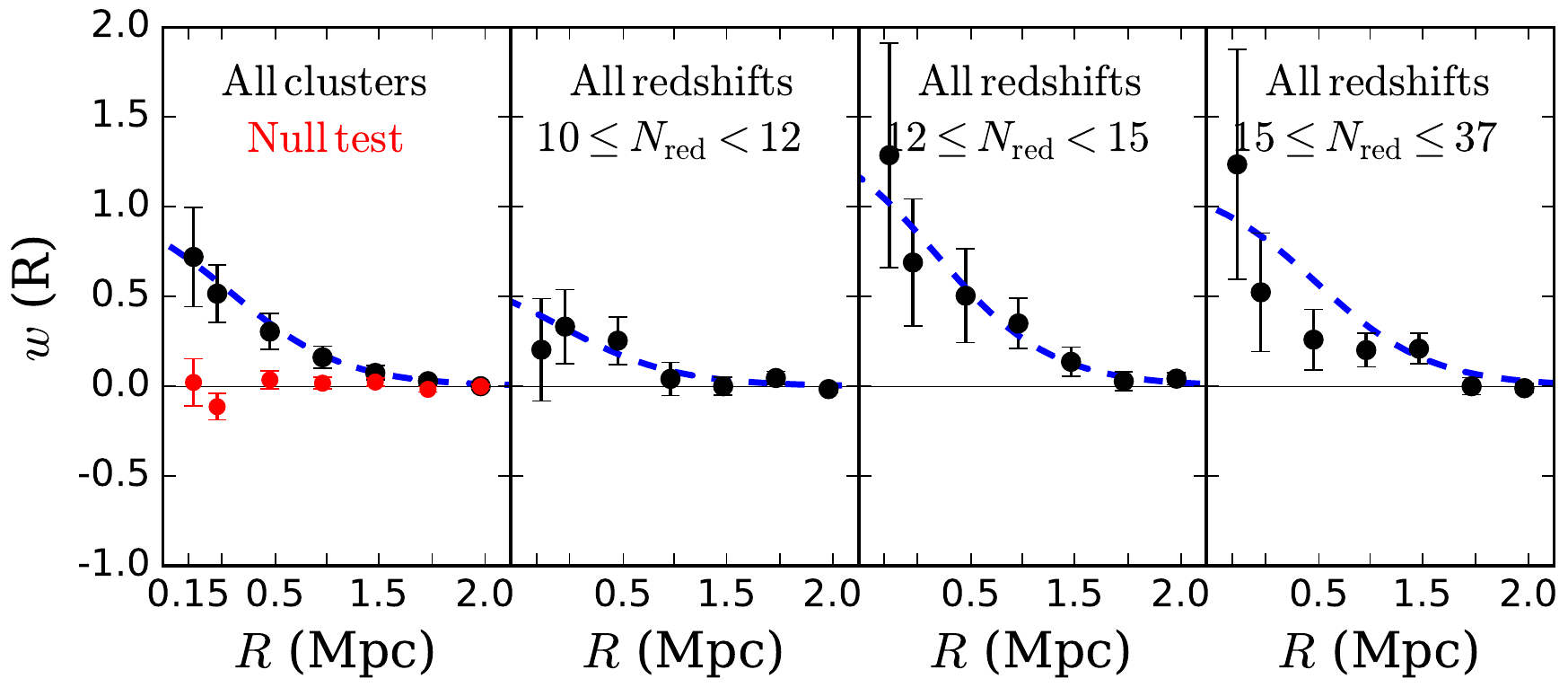}
 \caption{Same as in Fig. \ref{fig:correlation_z}, but with the cluster samples binned in richness instead of redshift. The red circles in the leftmost panel show the cross-correlation signal measured at random lens positions in our fields.}
 \label{fig:correlation_richness}
\end{figure*}

\begin{figure*}
 \centering
\includegraphics[width=138mm]{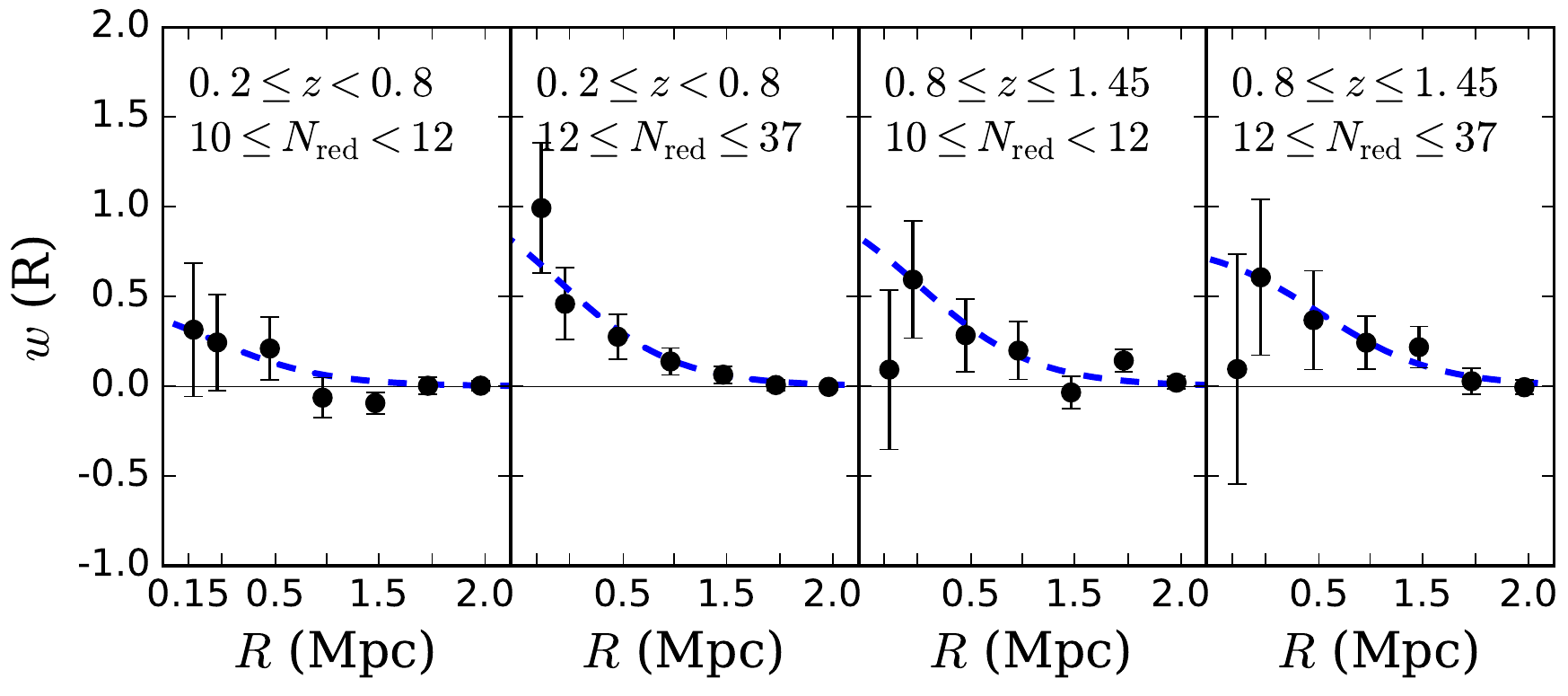}
 \caption{Same as in Fig. \ref{fig:correlation_z}, but with the cluster samples binned in richness and redshift as indicated in each panel.}
  \label{fig:correlation_z_Nred}
\end{figure*}

\begin{figure}
 \centering
\includegraphics[width=92mm]{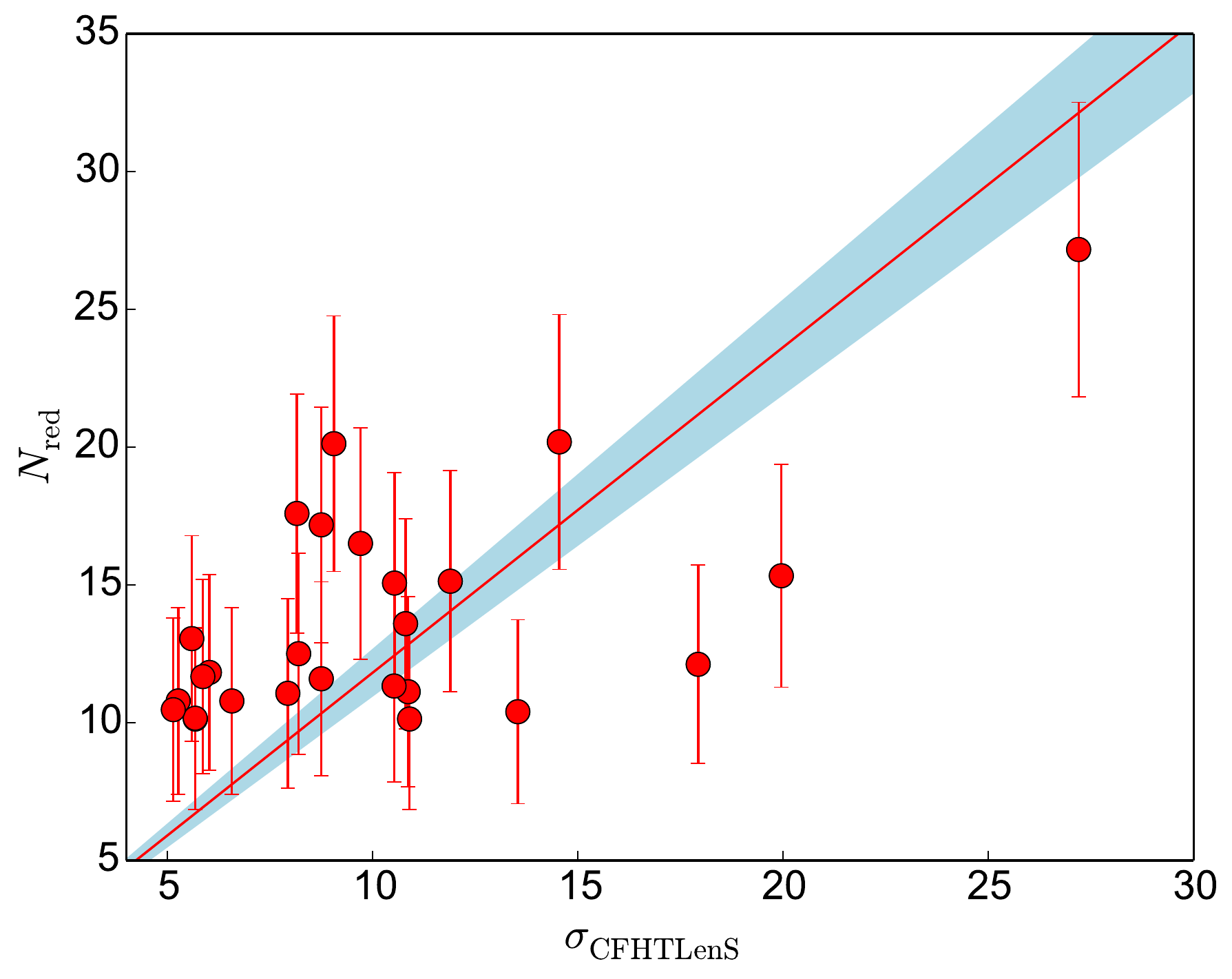}
 \caption{Richness proxies for the galaxy cluster candidates common to CFHTLenS and SpARCS. The best fit linear relation of the form $y = a x$ between the two quantities is shown by the continuous red line, while the associated fitting $1\sigma$ uncertainties are shown by the shaded region.}
 \label{fig:CFHTlens_Sparcs}
\end{figure}

\begin{figure}
 \centering
\includegraphics[width=91mm]{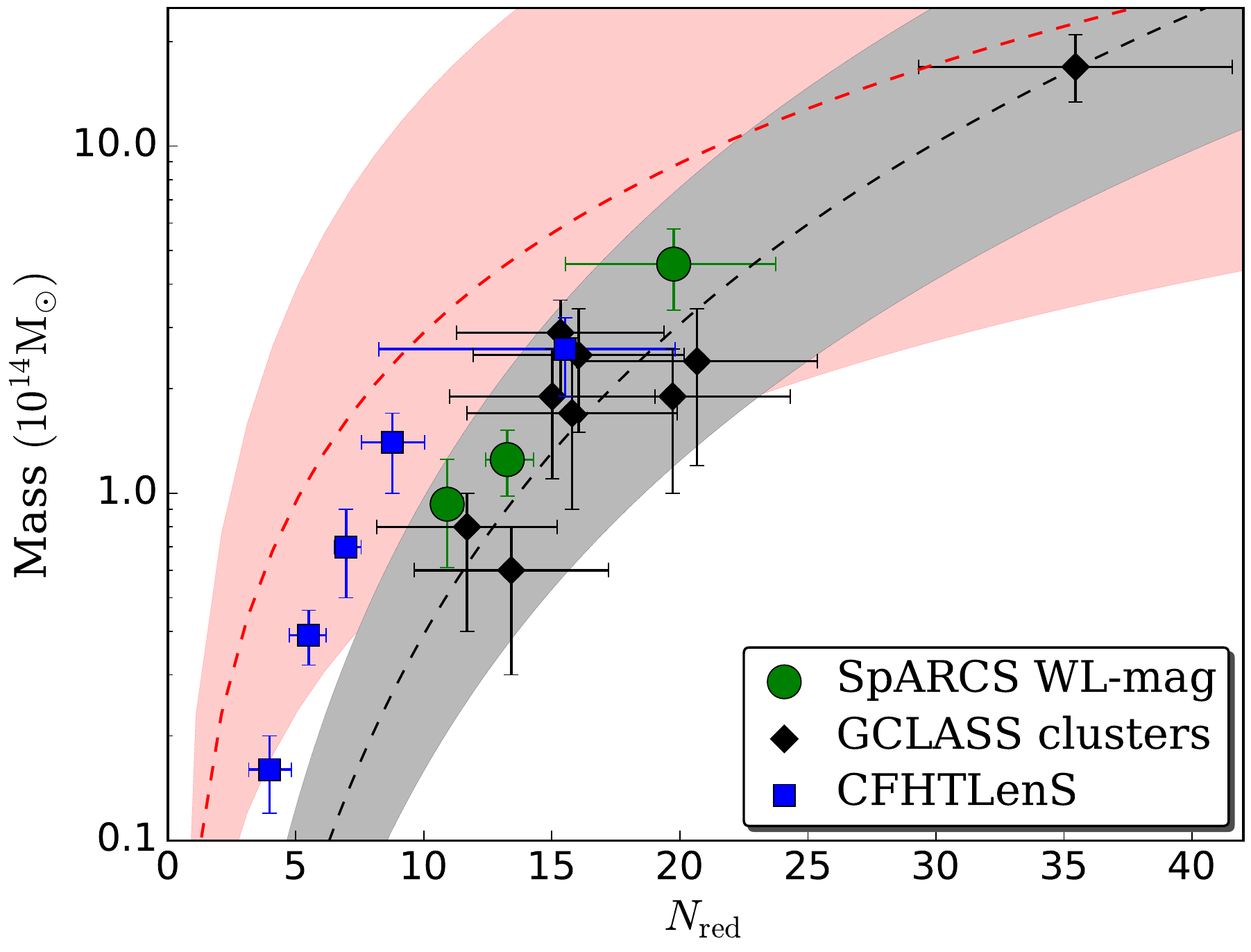}
 \caption{Mass and richness for the SpARCS cluster samples (green filled circles) compared to the the CFHTLenS analysis results \citep{Ford2014}, where the corresponding values are transformed into the $N_{\mathrm{red}}$ parameter as described in the text (blue filled squares). The errors of the CFHTLenS data points have the fit uncertainty from Fig. \ref{fig:CFHTlens_Sparcs} propagated as well. The red dashed line shows the mass-richness relation from Eq. \ref{eq:muzzin_mass_richness_relation}, with its uncertainty represented by the red shaded region, while the GCLASS-based mass-richness relation (Eq. \ref{eq:GCLASS_mass_richness_relation}) is shown by the black dashed line and respectively the shaded grey area.}
 \label{fig:mass_richness_relation}
\end{figure}

\begin{figure}
 \centering
\includegraphics[width=91mm]{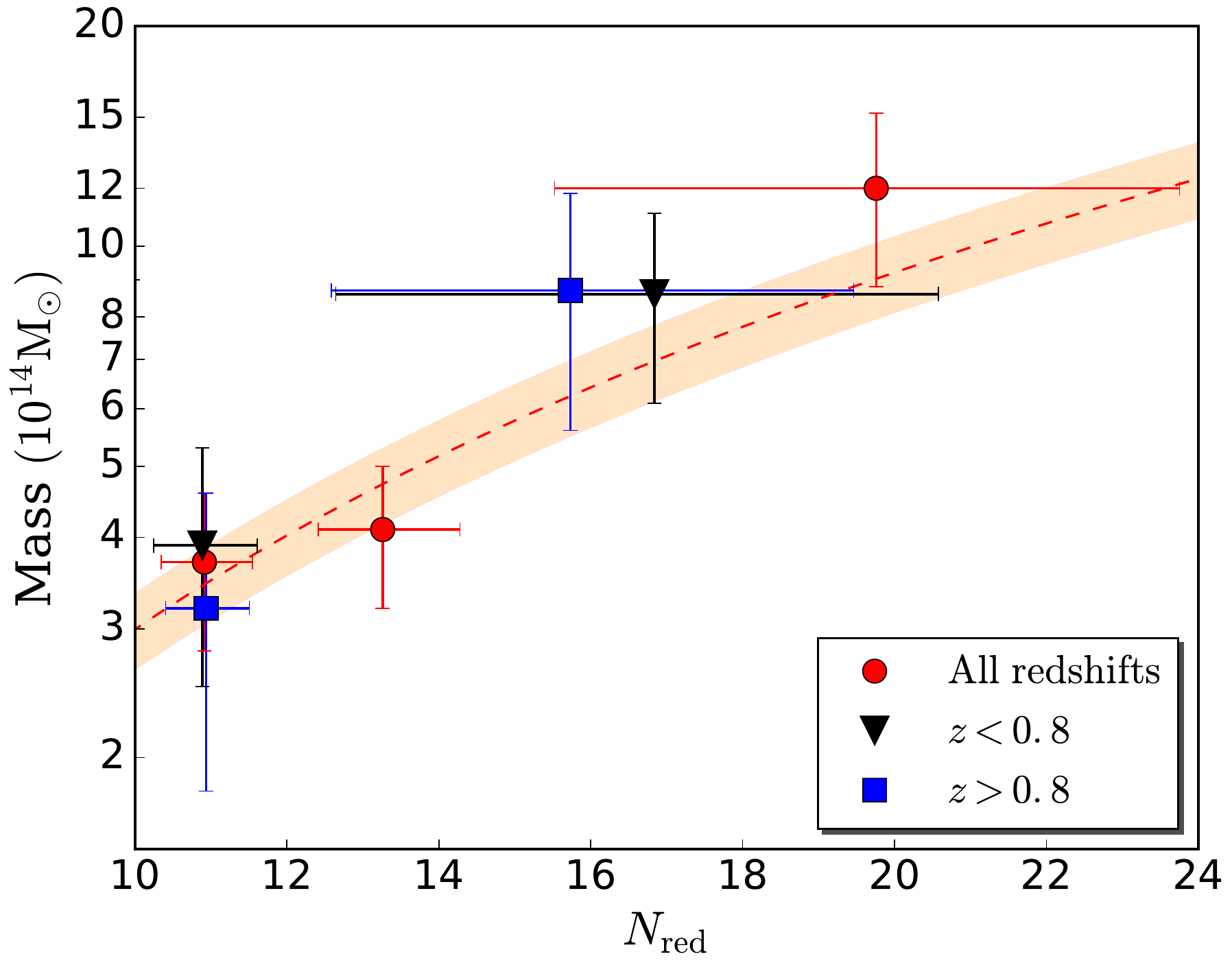}
 \caption{Mass-richness relation for the SpARCS cluster samples as a function of redshift. The dotted line and shaded region represent the best-fit mass-richness relation amplitude for all $z$, while the associated fit uncertainty is shown by the shaded region. The errorbars on the x-axis represent the ranges in $N_{\mathrm{red}}$ delimited by the 16$^{\mathrm{th}}$ and 84$^{\mathrm{th}}$ percentile for each sample.}
 \label{fig:mass_richness_relation_with_Nred}
\end{figure}

We split the cluster sample in several redshift bins, as can be seen in Fig. \ref{fig:correlation_z}. The uncertainty estimate on $w(\textrm{R})$ was computed by jackknife resampling over the measurements for all clusters included in the respective bin, while the lines show the best-fit model to the data. The redshift $z=0.8$ was chosen as marking the transition between the low and high-redshift samples based on the number of clusters available in each bin, with the main catalogue roughly split in half at this value. 

The measurement was carried out also on cluster samples binned in richness (see Fig. \ref{fig:correlation_richness}), with one sample containing all clusters with $10 \leq N_{\mathrm{red}} \leq 12$ and the other containing only the richest clusters in the sample, with $12 \leq N_{\mathrm{red}} \leq 37$. The first panel of Fig. \ref{fig:correlation_richness}) also shows the signal measured for large number of mock lenses situated at random positions in our survey. We measured the signal in various bins of redshift and richness of different widths, which can be seen in Fig. \ref{fig:correlation_z_Nred}. This particular binning was chosen in order to maximise the expected signal-to-noise (see Sec. \ref{Section:signal_to_noise}) by taking into account both the lensing efficiency as a function of angular diameter distance and keeping a roughly constant number of clusters below which the measurement errors become too large. Experimenting with our data set and bins of various widths, we found that in order to keep the measurement uncertainties to an acceptable level, a number of at least 50 clusters per bin is desirable.

 In all richness and redshift bins we measured the indicative signature of magnification with detection significances between 2.6$\sigma$ and 5.5$\sigma$. 
 
 It is difficult to compare our results for the mass-richness relation directly with other studies. $N_{\mathrm{red}}$, the richness proxy that we use, is not defined the same way as other richness estimates, and other studies will obtain different scaling relations depending on this particular choice, as well as on the details of the cluster detection algorithms. The uncertainties in estimating the richness parameter $N_{\mathrm{red}}$ were not propagated further in our analysis. 
 Fortunately, the CFHTLenS survey partly overlaps with the SpARCS area. The CFHTLenS \citep{Heymans2012,Erben2013} cluster catalogue is based on employing the 3D-Matched-Filter cluster-finder of \cite{Milkeraitis2010}, with cluster candidates spanning a wide range of masses ($10^{13} - 10^{15}$\,M$_\sun$) and redshifts ($0.2<z<0.9$). We found a number of 26 clusters with a high detection significance common to both catalogues, at similar redshifts and with a maximum separation smaller than $60\arcsec$. $\sigma_{\mathrm{CFHTLenS}}$ represents the significance of the likelihood peak relative to the background signal; \cite[see Eq. 5 of][for a detailed explanation]{Milkeraitis2010}. 
 
 \cite{Muzzin2007b} and \cite{Ford2014} utilize different definitions for richness and therefore we transformed the latter into the $N_{\mathrm{red}}$ parameter. To do so, we  fitted a linear relation that goes through the origin to the richness proxies used in the two studies, $N_{\mathrm{red}}$ SpARCS and  $\sigma_{\mathrm{CFHTLenS}}$, which can be seen in Fig. \ref{fig:CFHTlens_Sparcs}. The fitting errors represented in Fig. \ref{fig:CFHTlens_Sparcs} by the shaded region were propagated into the x-range errorbars for the CFHTLenS data points. Since the mass measurement results in \cite{Ford2014} are given as a function of $N_{200}$ instead of $\sigma_{\mathrm{CFHTLenS}}$, we included an additional conversion between these two parameters, which provides the result plotted in Fig. \ref{fig:mass_richness_relation}. We chose to first find the relation between the CFHTLenS $\sigma$ and  $N_{\mathrm{red}}$ because they are similar richness estimates and scale well together, unlike $N_{200}$ and $N_{\mathrm{red}}$. There are a number of caveats to this comparison, such as additional uncertainties and systematic biases that we do not take into account. Among these, probably most important are the fact that only a very small number of clusters is common to the two studies, which could introduce selection bias effects, as well as the rather large cut-off in separation when matching the two catalogues, which could mean that some clusters are erroneously matched. 
 
 Figure \ref{fig:mass_richness_relation} shows the mass and richness derived from our measurements and results from the methodologically-similar study based on CFHTLenS data, as well as the GCLASS clusters.
 
 Figure \ref{fig:mass_richness_relation_with_Nred} shows the mass-richness relation for the most relevant  $N_{\mathrm{red}}$ and redshift bins, as well as the best-fit mass-richness relation and its uncertainty. The horizontal error bars for the SpARCS measurements represent the ranges in $N_{\mathrm{red}}$ delimited by the 16$^{\mathrm{th}}$ and 84$^{\mathrm{th}}$ percentile for each sample. We do not observe a clear trend of the mass-richness relation with redshift.

  We selected the six GCLASS clusters that are within the area covered by the CFHT pointings for measuring the magnification signal. However, the measurement appeared to be too noisy for a meaningful interpretation, which can be explained by insufficient number density of the $u$-dropouts. 
  
\section{Discussion} \label{Section:Discussion}

We discuss below some of the most important effects that we are not modelling and which could potentially impact our results and other measurements based on similar techniques.

Simulations by \cite{Hildebrandt2009} for a significantly deeper data set show that by using the cuts in magnitude that we also used, the contamination from stars and low-z interlopers is below 10\% in each 0.5 magnitude interval. Since our data are shallower than the one for which  \cite{Hildebrandt2009} assessed the contamination, this means that the interloper proportion might be higher in our sample, therefore contributing to a dilution of the signal strength (if more stars are added into the sample) or to an additional, unwanted, cross-correlation signal between low-redshift interlopers and the low-redshift cluster sample. To minimize the probability that there is a physical cross-correlation between low-redshift contaminants in our LBG candidates sample and low-redshift galaxy clusters, we also tested a sample of cluster candidates with $0.3\leq z<0.8$. This however resulted in practically identical estimates with those given by the clusters in the $0.2\leq z<0.8$ bin, since there are very few cluster candidates with such a low redshift. Therefore low redshift contamination most likely does not play an important role for our measurements.

\begin{table*}\setlength\extrarowheight{2pt} 
\centering
    \caption{Results of the mass measurements for cluster samples binned in redshift and richness (see Fig. \ref{fig:correlation_z} and Fig. \ref{fig:correlation_richness}).}  
    \begin{centering}
    \begin{tabular}{ccccccccc}
      \hline
      \hline
     Sample description & Redshift range &  $N_{\mathrm{red}}$ range & Average $N_{\mathrm{red}}$ & Number of clusters & $\langle M_{200} \rangle $  & $\chi^2_{\mathrm{red}}$ & S/N & $a$ \\ 
     &$(z)$ & & & & $ \left( 10^{14}\,\textrm{M}_{\sun}\right)$  & & &\\
      \hline
\multirow{ 5}{*}{$z$-bins}     
       &$ 0.2 - 1.4 $	&  10-37 & 13.9 & 287 & 1.28  $^{+0.23}_{-0.21}$    & 0.8 & 5.5 & $1.21^{+0.22}_{-0.20}$ \\
       &$ 0.2 - 0.5 $	&  10-37 & 14.5 & 71  & 1.56  $^{+0.30}_{-0.31}$    & 1.4 & 4.6 & $1.30^{+0.25}_{-0.26}$ \\
       &$ 0.5 - 0.8 $	&  10-37 & 14.4 & 83  & 1.31  $^{+0.31}_{-0.27}$    & 2.2 & 3.5 & $1.11^{+0.26}_{-0.23}$ \\
       &$ 0.8 - 1.0	$	&  10-37 & 13.3 & 67  & 1.19  $^{+0.27}_{-0.33}$    & 1.9 & 2.6 & $1.28^{+0.29}_{-0.36}$ \\
 	   &$ 1.0 - 1.4	$	&  10-37 & 12.8 & 66  & 1.03  $^{+0.24}_{-0.28}$    & 2.3 & 4.1 & $1.24^{+0.29}_{-0.34}$ \\
       \hline
\multirow{ 3}{*}{$N_{\mathrm{red}}$-bins}       
       &$ 0.2 - 1.4 $	&  10-12 & 10.9 & 134 & 0.93  $^{+0.32}_{-0.21}$    & 1.8 & 3.2 & $1.81^{+0.62}_{-0.41}$ \\
       &$ 0.2 - 1.4 $	&  12-15 & 13.3 & 77  & 1.25  $^{+0.27}_{-0.23}$    & 2.5 & 4.4 & $1.35^{+0.29}_{-0.25}$ \\
       &$ 0.2 - 1.4	$	&  15-37 & 19.5 & 75  & 4.57  $^{+1.25}_{-1.16}$    & 3.2 & 4.0 & $1.53^{+0.42}_{-0.39}$ \\
       \hline
\multirow{ 2}{*}{high-$z$-bins} 	   
       &$ 0.8 - 1.4	$	&  10-12 & 10.9 & 67  & 0.87 $^{+0.20}_{-0.22}$     & 2.6 & 3.1 & $1.69^{+0.39}_{-0.43}$ \\
       &$ 0.8 - 1.4 $	&  12-37 & 15.8 & 66  & 3.44 $^{+1.14}_{-1.19}$     & 2.4 & 3.0 & $2.22^{+0.74}_{-0.77}$ \\
       \hline
\multirow{ 2}{*}{low-$z$-bins} 	   
       &$ 0.2 - 0.8	$	&  10-12 & 10.9 & 61  & 1.20 $^{+0.41}_{-0.38}$     & 1.6 & 2.8 & $2.33^{+0.80}_{-0.74}$ \\
       &$ 0.2 - 0.8 $	&  12-37 & 16.9 & 93  & 3.29 $^{+1.08}_{-1.11}$     & 1.7 & 4.4 & $1.74^{+0.54}_{-0.59}$ \\
      \hline
    \end{tabular}
    \end{centering}
    \label{mass_estimates}
\end{table*} 

Besides the effective masking described in Sect. \ref{Section:masking_correction}, cluster galaxies might have another critical influence on the detection of LBG candidates. When $u$-dropouts are in the angular proximity of cluster member galaxies, the measurement of their colours could possibly also be affected. In effect, this could shift the entire population of LBGs that are close to cluster galaxies across colour space, either by increasing magnitude measurement errors (therefore including more fake candidates and rejecting real ones, thus increasing the overall noise of the measurement) or by consistently shifting LBGs in the colour space as a function of the overall galaxy cluster colours, thus creating a redshift-dependent effect.

Additionally, the method used by {\tt SExtractor} for background subtraction results in under- and over-subtraction at various distances from the field galaxies, especially those of sizes closest to the smoothing kernel used to estimate the background map. The procedures {\tt SExtractor} uses to separate partly blended objects is also of relevance and must be investigated in a comprehensive manner. Understanding these effects requires detailed simulations which go beyond the scope of this study and will be left to future research.

Increasing the total area by including the high quality DECam data available for the southern SpARCS fields and modelling these additional effects could be of great assistance in decreasing the noise and increasing especially the significance and precision of the high-redshift cross-correlation measurements. This would enable us to even more accurately calibrate the mass-richness relation at $z \geq 0.8$ and to use similar magnification-based methods for studying large samples of galaxy clusters with more precision and accuracy.

In this paper we discuss many different possible systematic errors in the measurement. We neglect several possible systematic effects on the modelling side that are usually addressed in the galaxy cluster literature, such as the effect of large-scale-structure along the line of sight, triaxiality of galaxy clusters, uncertainty in the $M$ - $c$ relation, etc. Given the size of our statistical errors we can be confident that such effects are subdominant at the moment and defer their treatment to future work.

\section{Summary and conclusions} \label{conclusions}

We used optically-selected LBGs to study a sample of galaxy cluster candidates by using the magnification bias induced by the weak gravitational lensing magnification effect. A total of 287 galaxy cluster candidates with a high detection significance were selected from the SpARCS catalogue, with redshifts ranging from 0.2 to 1.4 and richness \citep[as defined by][]{Muzzin2008} between 10 and 37. Using the Lyman-break technique on deep $ugriz$ optical data from CFHT, we selected a background sample of 16\,242 objects with a magnitude range of $23.0\leq r \leq 24.5$ and situated at a redshift of $z \approx$ 3.1, offering both a sufficient surface number density and good lensing efficiency. We cross-correlated the positions of the galaxy cluster candidates and the LBGs, using an external LBG luminosity function to calibrate our measurement. We fitted a composite NFW halo model that takes into account the richness and redshift ranges of the cluster candidate sample, as well as modelling contributions from the two-halo term, miscentering and low redshift contamination.

We report a $5.5 \sigma$ detection significance for the weak lensing magnification signal $w$(R), measured for the entire cluster dataset. We found an average halo mass for the cluster sample with $0.2 \leq z \leq 1.4$ of $M_{200}=1.28^{+0.23}_{-0.21} \times 10^{14}$\,M$_\sun$. The cluster catalogue was divided in various richness and/or redshift bins, with the mass and normalization of the mass-richness relation parameter $a$ estimates for each bin presented in Table \ref{mass_estimates}. As we only fitted the amplitude $a$ of a specific mass-richness relation, it is important that we use a representative mass-richness relation for the data. Ideally, we would fit both the normalisation and the slope, thus reducing the dependence on the exact shape of the mass-richness relation at the basis of the model fitting.

Although the contamination of the cluster catalogue with spurious detections is not known precisely, our results indicate that optical-IR selection of clusters does in fact select real, massive over-densities even at very high redshift ($z\geq1$) and is a promising and efficient method for selecting large samples of such objects with a relatively low observational effort. The significance of the measurement for clusters at high redshift ($z\geq1$) is a remarkable 4.1 $\sigma$, thus further strengthening the case for using weak lensing magnification methods to calibrate the mass-richness relation for large samples of high redshift galaxy clusters. 

Even if the mass-richness scaling relation is directly applicable only to the cluster sample from which it was obtained, the ease of measuring richness for any optical survey makes richness an important quantity to measure. A meta-study that would compare how different richness proxies relate to each other could provide a bridge for having more direct comparisons between studies. 

Albeit the signal-to-noise ratio for mass measurements obtained using weak lensing shear methods is higher in general, considerable improvements in using magnification as a complementary method are achieved. Additionally, magnification probes the surface mass density of the lens directly, while shear measures the differential mass density, thus the combination of these two methods is able to break the lens mass-sheet degeneracy. \cite{Rozo2010} have shown that survey-independent statistical gains of the order of 40\% - 50\% can be obtained by using the two types of measurements together.

Next generation surveys, such as Kilo Degree Survey \citep[KiDS,][]{KiDS}, Euclid \citep{Laureijs2011}, Subaru Hyper Suprime-Cam \citep{SHSC}, Large Synoptic Survey Telescope \citep[LSST,][]{LSST}, Dark Energy Survey \citep[DES,][]{DES}, etc. will be expected to be able to take full advantage of the large areas covered, large number of background sources and excellent redshift estimates in order to use the combined strengths of weak lensing shear and magnification for a large class of cosmological and weak lensing science problems. Having a well understood and appropriately calibrated mass-richness relation before LSST starts providing large data sets is critical for enabling the measurement of $M_{200}$ for an unprecedented large number of cluster from imaging data alone, thus enabling very accurate cosmological studies, such as greatly strengthening constrains on the dark energy equation of state from the cluster mass function.

\section*{Acknowledgements}

The authors A.T., H.H., M.T. and C.B.M. have been supported by the DFG Emmy Noether grant Hi 1495/2-1.

R.D. gratefully acknowledges the support provided by the BASAL Center for Astrophysics and Associated Technologies (CATA).

J.N. is supported by Universdad Andres Bello internal grant number DI-18-17/RG.

G.W. acknowledges financial support for this work from NSF grant AST-1517863 and from NASA through programmes GO-13306, GO-13677, GO-13747 \& GO-13845/14327 from the Space Telescope Science Institute, which is operated by AURA, Inc., under NASA contract NAS 5-26555.

This work is based in part on observations obtained with MegaCam, a joint project of CFHT and CEA/IRFU, at the CFHT which is operated by the National Research Council (NRC) of Canada, the Institut National des Sciences de l'Univers of the Centre National de la Recherche Scientifique (CNRS) of France, and the University of Hawaii. This research used the facilities of the Canadian Astronomy Data Centre operated by the National Research Council of Canada with the support of the Canadian Space Agency. CFHTLenS data processing was made possible thanks to significant computing support from the NSERC Research Tools and Instruments grant programme.

This work is also partly based on observations made with the Spitzer Space Telescope, which is operated by the Jet Propulsion Laboratory, California Institute of Technology under a contract with NASA.

We would like to thank the anonymous referee for comments that have significantly improved the clarity of the paper.
 
\bibliographystyle{aa}
\bibliography{references.bib}

\begin{thebibliography}{77}
\expandafter\ifx\csname natexlab\endcsname\relax\def\natexlab#1{#1}\fi

\bibitem[{{Ahn} {et~al.}(2014){Ahn}, {Alexandroff}, {Allende Prieto}, {Anders},
  {Anderson}, {Anderton}, {Andrews}, {Aubourg}, {Bailey}, {Bastien}, \&
  et~al.}]{SDSSDR10}
{Ahn}, C.~P., {Alexandroff}, R., {Allende Prieto}, C., {et~al.} 2014, \apjs,
  211, 17

\bibitem[{{Allen} {et~al.}(2011){Allen}, {Evrard}, \& {Mantz}}]{Allen2011}
{Allen}, S.~W., {Evrard}, A.~E., \& {Mantz}, A.~B. 2011, \araa, 49, 409

\bibitem[{{Applegate} {et~al.}(2014){Applegate}, {von der Linden}, {Kelly},
  {Allen}, {Allen}, {Burchat}, {Burke}, {Ebeling}, {Mantz}, \&
  {Morris}}]{Applegate2014}
{Applegate}, D.~E., {von der Linden}, A., {Kelly}, P.~L., {et~al.} 2014,
  \mnras, 439, 48

\bibitem[{{Bartelmann}(1996)}]{Bartelmann1996}
{Bartelmann}, M. 1996, \aap, 313, 697

\bibitem[{{Bartelmann} \& {Schneider}(2001)}]{Bartelmann2001}
{Bartelmann}, M. \& {Schneider}, P. 2001, \physrep, 340, 291

\bibitem[{{Ben{\'{\i}}tez}(2000)}]{Benitez2000}
{Ben{\'{\i}}tez}, N. 2000, \apj, 536, 571

\bibitem[{{Bertin}(2006)}]{Bertin2006}
{Bertin}, E. 2006, in Astronomical Society of the Pacific Conference Series,
  Vol. 351, Astronomical Data Analysis Software and Systems XV, ed.
  C.~{Gabriel}, C.~{Arviset}, D.~{Ponz}, \& S.~{Enrique}, 112

\bibitem[{{Bertin} \& {Arnouts}(1996)}]{Bertin1996}
{Bertin}, E. \& {Arnouts}, S. 1996, \aaps, 117, 393

\bibitem[{{Bertin} {et~al.}(2002){Bertin}, {Mellier}, {Radovich}, {Missonnier},
  {Didelon}, \& {Morin}}]{Bertin2002}
{Bertin}, E., {Mellier}, Y., {Radovich}, M., {et~al.} 2002, in Astronomical
  Society of the Pacific Conference Series, Vol. 281, Astronomical Data
  Analysis Software and Systems XI, ed. D.~A. {Bohlender}, D.~{Durand}, \&
  T.~H. {Handley}, 228

\bibitem[{{Blindert} {et~al.}(2004){Blindert}, {Yee}, {Gladders}, {Ellingson},
  \& {Golding}}]{Blindert2004}
{Blindert}, K., {Yee}, H.~K.~C., {Gladders}, M.~D., {Ellingson}, E., \&
  {Golding}, J. 2004, in Bulletin of the American Astronomical Society,
  Vol.~36, American Astronomical Society Meeting Abstracts, 1609

\bibitem[{{de Jong} {et~al.}(2013){de Jong}, {Kuijken}, {Applegate}, {Begeman},
  {Belikov}, {Blake}, {Bout}, {Boxhoorn}, {Buddelmeijer}, {Buddendiek},
  {Cacciato}, {Capaccioli}, {Choi}, {Cordes}, {Covone}, {Dall'Ora}, {Edge},
  {Erben}, {Franse}, {Getman}, {Grado}, {Harnois-Deraps}, {Helmich},
  {Herbonnet}, {Heymans}, {Hildebrandt}, {Hoekstra}, {Huang}, {Irisarri},
  {Joachimi}, {K{\"o}hlinger}, {Kitching}, {La Barbera}, {Lacerda},
  {McFarland}, {Miller}, {Nakajima}, {Napolitano}, {Paolillo}, {Peacock},
  {Pila-Diez}, {Puddu}, {Radovich}, {Rifatto}, {Schneider}, {Schrabback},
  {Sifon}, {Sikkema}, {Simon}, {Sutherland}, {Tudorica}, {Valentijn}, {van der
  Burg}, {van Uitert}, {van Waerbeke}, {Velander}, {Kleijn}, {Viola}, \&
  {Vriend}}]{KiDS}
{de Jong}, J.~T.~A., {Kuijken}, K., {Applegate}, D., {et~al.} 2013, The
  Messenger, 154, 44

\bibitem[{{Dietrich} {et~al.}(2007){Dietrich}, {Erben}, {Lamer}, {Schneider},
  {Schwope}, {Hartlap}, \& {Maturi}}]{Dietrich2007}
{Dietrich}, J.~P., {Erben}, T., {Lamer}, G., {et~al.} 2007, \aap, 470, 821

\bibitem[{{Duffy} {et~al.}(2008){Duffy}, {Schaye}, {Kay}, \& {Dalla
  Vecchia}}]{Duffy2008}
{Duffy}, A.~R., {Schaye}, J., {Kay}, S.~T., \& {Dalla Vecchia}, C. 2008,
  \mnras, 390, L64

\bibitem[{{Erben} {et~al.}(2009){Erben}, {Hildebrandt}, {Lerchster}, {Hudelot},
  {Benjamin}, {van Waerbeke}, {Schrabback}, {Brimioulle}, {Cordes}, {Dietrich},
  {Holhjem}, {Schirmer}, \& {Schneider}}]{Erben2009}
{Erben}, T., {Hildebrandt}, H., {Lerchster}, M., {et~al.} 2009, \aap, 493, 1197

\bibitem[{{Erben} {et~al.}(2013){Erben}, {Hildebrandt}, {Miller}, {van
  Waerbeke}, {Heymans}, {Hoekstra}, {Kitching}, {Mellier}, {Benjamin}, {Blake},
  {Bonnett}, {Cordes}, {Coupon}, {Fu}, {Gavazzi}, {Gillis}, {Grocutt}, {Gwyn},
  {Holhjem}, {Hudson}, {Kilbinger}, {Kuijken}, {Milkeraitis}, {Rowe},
  {Schrabback}, {Semboloni}, {Simon}, {Smit}, {Toader}, {Vafaei}, {van Uitert},
  \& {Velander}}]{Erben2013}
{Erben}, T., {Hildebrandt}, H., {Miller}, L., {et~al.} 2013, \mnras, 433, 2545

\bibitem[{{Fazio} {et~al.}(2004){Fazio}, {Hora}, {Allen}, {Ashby}, {Barmby},
  {Deutsch}, {Huang}, {Kleiner}, {Marengo}, {Megeath}, {Melnick}, {Pahre},
  {Patten}, {Polizotti}, {Smith}, {Taylor}, {Wang}, {Willner}, {Hoffmann},
  {Pipher}, {Forrest}, {McMurty}, {McCreight}, {McKelvey}, {McMurray}, {Koch},
  {Moseley}, {Arendt}, {Mentzell}, {Marx}, {Losch}, {Mayman}, {Eichhorn},
  {Krebs}, {Jhabvala}, {Gezari}, {Fixsen}, {Flores}, {Shakoorzadeh}, {Jungo},
  {Hakun}, {Workman}, {Karpati}, {Kichak}, {Whitley}, {Mann}, {Tollestrup},
  {Eisenhardt}, {Stern}, {Gorjian}, {Bhattacharya}, {Carey}, {Nelson},
  {Glaccum}, {Lacy}, {Lowrance}, {Laine}, {Reach}, {Stauffer}, {Surace},
  {Wilson}, {Wright}, {Hoffman}, {Domingo}, \& {Cohen}}]{Fazio2004}
{Fazio}, G.~G., {Hora}, J.~L., {Allen}, L.~E., {et~al.} 2004, \apjs, 154, 10

\bibitem[{{Ford} {et~al.}(2014){Ford}, {Hildebrandt}, {Van Waerbeke}, {Erben},
  {Laigle}, {Milkeraitis}, \& {Morrison}}]{Ford2014}
{Ford}, J., {Hildebrandt}, H., {Van Waerbeke}, L., {et~al.} 2014, \mnras, 439,
  3755

\bibitem[{{Ford} {et~al.}(2012){Ford}, {Hildebrandt}, {Van Waerbeke},
  {Leauthaud}, {Capak}, {Finoguenov}, {Tanaka}, {George}, \&
  {Rhodes}}]{Ford2012}
{Ford}, J., {Hildebrandt}, H., {Van Waerbeke}, L., {et~al.} 2012, \apj, 754,
  143

\bibitem[{{Giavalisco}(2002)}]{Giavalisco2002}
{Giavalisco}, M. 2002, \araa, 40, 579

\bibitem[{{Giavalisco} {et~al.}(1996){Giavalisco}, {Steidel}, \&
  {Macchetto}}]{Giavalisco1996b}
{Giavalisco}, M., {Steidel}, C.~C., \& {Macchetto}, F.~D. 1996, \apj, 470, 189

\bibitem[{{Gilbank} {et~al.}(2004){Gilbank}, {Bower}, {Castander}, \&
  {Ziegler}}]{Gilbank2004}
{Gilbank}, D.~G., {Bower}, R.~G., {Castander}, F.~J., \& {Ziegler}, B.~L. 2004,
  \mnras, 348, 551

\bibitem[{{Gilbank} {et~al.}(2007){Gilbank}, {Yee}, {Ellingson}, {Gladders},
  {Barrientos}, \& {Blindert}}]{Gilbank2007}
{Gilbank}, D.~G., {Yee}, H.~K.~C., {Ellingson}, E., {et~al.} 2007, \aj, 134,
  282

\bibitem[{{Giodini} {et~al.}(2013){Giodini}, {Lovisari}, {Pointecouteau},
  {Ettori}, {Reiprich}, \& {Hoekstra}}]{Giodini2013}
{Giodini}, S., {Lovisari}, L., {Pointecouteau}, E., {et~al.} 2013, \ssr, 177,
  247

\bibitem[{{Gladders} \& {Yee}(2000)}]{Gladders2000}
{Gladders}, M.~D. \& {Yee}, H.~K.~C. 2000, \aj, 120, 2148

\bibitem[{{Gladders} \& {Yee}(2005)}]{Gladders2005}
{Gladders}, M.~D. \& {Yee}, H.~K.~C. 2005, \apjs, 157, 1

\bibitem[{{Gruen} {et~al.}(2014){Gruen}, {Seitz}, {Brimioulle}, {Kosyra},
  {Koppenhoefer}, {Lee}, {Bender}, {Riffeser}, {Eichner}, {Weidinger}, \&
  {Bierschenk}}]{Gruen2014}
{Gruen}, D., {Seitz}, S., {Brimioulle}, F., {et~al.} 2014, \mnras, 442, 1507

\bibitem[{{Hartlap} {et~al.}(2007){Hartlap}, {Simon}, \&
  {Schneider}}]{Hartlap2007}
{Hartlap}, J., {Simon}, P., \& {Schneider}, P. 2007, \aap, 464, 399

\bibitem[{{Heymans} {et~al.}(2012){Heymans}, {Van Waerbeke}, {Miller}, {Erben},
  {Hildebrandt}, {Hoekstra}, {Kitching}, {Mellier}, {Simon}, {Bonnett},
  {Coupon}, {Fu}, {Harnois D{\'e}raps}, {Hudson}, {Kilbinger}, {Kuijken},
  {Rowe}, {Schrabback}, {Semboloni}, {van Uitert}, {Vafaei}, \&
  {Velander}}]{Heymans2012}
{Heymans}, C., {Van Waerbeke}, L., {Miller}, L., {et~al.} 2012, \mnras, 427,
  146

\bibitem[{{Hildebrandt} {et~al.}(2012){Hildebrandt}, {Erben}, {Kuijken}, {van
  Waerbeke}, {Heymans}, {Coupon}, {Benjamin}, {Bonnett}, {Fu}, {Hoekstra},
  {Kitching}, {Mellier}, {Miller}, {Velander}, {Hudson}, {Rowe}, {Schrabback},
  {Semboloni}, \& {Ben{\'{\i}}tez}}]{Hildebrandt2012}
{Hildebrandt}, H., {Erben}, T., {Kuijken}, K., {et~al.} 2012, \mnras, 421, 2355

\bibitem[{{Hildebrandt} {et~al.}(2011){Hildebrandt}, {Muzzin}, {Erben},
  {Hoekstra}, {Kuijken}, {Surace}, {van Waerbeke}, {Wilson}, \&
  {Yee}}]{Hildebrandt2011}
{Hildebrandt}, H., {Muzzin}, A., {Erben}, T., {et~al.} 2011, \apjl, 733, L30

\bibitem[{{Hildebrandt} {et~al.}(2009){Hildebrandt}, {Pielorz}, {Erben}, {van
  Waerbeke}, {Simon}, \& {Capak}}]{Hildebrandt2009}
{Hildebrandt}, H., {Pielorz}, J., {Erben}, T., {et~al.} 2009, \aap, 498, 725

\bibitem[{{Hildebrandt} {et~al.}(2013){Hildebrandt}, {van Waerbeke}, {Scott},
  {B{\'e}thermin}, {Bock}, {Clements}, {Conley}, {Cooray}, {Dunlop}, {Eales},
  {Erben}, {Farrah}, {Franceschini}, {Glenn}, {Halpern}, {Heinis}, {Ivison},
  {Marsden}, {Oliver}, {Page}, {P{\'e}rez-Fournon}, {Smith}, {Rowan-Robinson},
  {Valtchanov}, {van der Burg}, {Vieira}, {Viero}, \& {Wang}}]{Hildebrandt2013}
{Hildebrandt}, H., {van Waerbeke}, L., {Scott}, D., {et~al.} 2013, \mnras, 429,
  3230

\bibitem[{{Hoekstra} {et~al.}(2015){Hoekstra}, {Herbonnet}, {Muzzin}, {Babul},
  {Mahdavi}, {Viola}, \& {Cacciato}}]{Hoekstra2015}
{Hoekstra}, H., {Herbonnet}, R., {Muzzin}, A., {et~al.} 2015, \mnras, 449, 685

\bibitem[{{Ivezic} {et~al.}(2008){Ivezic}, {Tyson}, {Abel}, {Acosta},
  {Allsman}, {AlSayyad}, {Anderson}, {Andrew}, {Angel}, {Angeli}, {Ansari},
  {Antilogus}, {Arndt}, {Astier}, {Aubourg}, {Axelrod}, {Bard}, {Barr},
  {Barrau}, {Bartlett}, {Bauman}, {Beaumont}, {Becker}, {Becla}, {Beldica},
  {Bellavia}, {Blanc}, {Blandford}, {Bloom}, {Bogart}, {Borne}, {Bosch},
  {Boutigny}, {Brandt}, {Brown}, {Bullock}, {Burchat}, {Burke}, {Cagnoli},
  {Calabrese}, {Chandrasekharan}, {Chesley}, {Cheu}, {Chiang}, {Claver},
  {Connolly}, {Cook}, {Cooray}, {Covey}, {Cribbs}, {Cui}, {Cutri}, {Daubard},
  {Daues}, {Delgado}, {Digel}, {Doherty}, {Dubois}, {Dubois-Felsmann},
  {Durech}, {Eracleous}, {Ferguson}, {Frank}, {Freemon}, {Gangler}, {Gawiser},
  {Geary}, {Gee}, {Geha}, {Gibson}, {Gilmore}, {Glanzman}, {Goodenow},
  {Gressler}, {Gris}, {Guyonnet}, {Hascall}, {Haupt}, {Hernandez}, {Hogan},
  {Huang}, {Huffer}, {Innes}, {Jacoby}, {Jain}, {Jee}, {Jernigan},
  {Jevremovic}, {Johns}, {Jones}, {Juramy-Gilles}, {Juric}, {Kahn}, {Kalirai},
  {Kallivayalil}, {Kalmbach}, {Kantor}, {Kasliwal}, {Kessler}, {Kirkby},
  {Knox}, {Kotov}, {Krabbendam}, {Krughoff}, {Kubanek}, {Kuczewski},
  {Kulkarni}, {Lambert}, {Le Guillou}, {Levine}, {Liang}, {Lim}, {Lintott},
  {Lupton}, {Mahabal}, {Marshall}, {Marshall}, {May}, {McKercher}, {Migliore},
  {Miller}, {Mills}, {Monet}, {Moniez}, {Neill}, {Nief}, {Nomerotski},
  {Nordby}, {O'Connor}, {Oliver}, {Olivier}, {Olsen}, {Ortiz}, {Owen}, {Pain},
  {Peterson}, {Petry}, {Pierfederici}, {Pietrowicz}, {Pike}, {Pinto}, {Plante},
  {Plate}, {Price}, {Prouza}, {Radeka}, {Rajagopal}, {Rasmussen}, {Regnault},
  {Ridgway}, {Ritz}, {Rosing}, {Roucelle}, {Rumore}, {Russo}, {Saha},
  {Sassolas}, {Schalk}, {Schindler}, {Schneider}, {Schumacher}, {Sebag},
  {Sembroski}, {Seppala}, {Shipsey}, {Silvestri}, {Smith}, {Smith}, {Strauss},
  {Stubbs}, {Sweeney}, {Szalay}, {Takacs}, {Thaler}, {Van Berg}, {Vanden Berk},
  {Vetter}, {Virieux}, {Xin}, {Walkowicz}, {Walter}, {Wang}, {Warner},
  {Willman}, {Wittman}, {Wolff}, {Wood-Vasey}, {Yoachim}, {Zhan}, \& {for the
  LSST Collaboration}}]{LSST}
{Ivezic}, Z., {Tyson}, J.~A., {Abel}, B., {et~al.} 2008, ArXiv e-prints,
  astro-ph/0805.2366

\bibitem[{{Johnston} {et~al.}(2007){Johnston}, {Sheldon}, {Wechsler}, {Rozo},
  {Koester}, {Frieman}, {McKay}, {Evrard}, {Becker}, \& {Annis}}]{Johnston2007}
{Johnston}, D.~E., {Sheldon}, E.~S., {Wechsler}, R.~H., {et~al.} 2007, ArXiv
  e-prints, astro-ph/0709.1159

\bibitem[{{Kuijken}(2008)}]{Kuijken2008}
{Kuijken}, K. 2008, \aap, 482, 1053

\bibitem[{{Laureijs} {et~al.}(2011){Laureijs}, {Amiaux}, {Arduini},
  {Augu{\`e}res}, {Brinchmann}, {Cole}, {Cropper}, {Dabin}, {Duvet}, {Ealet},
  \& et~al.}]{Laureijs2011}
{Laureijs}, R., {Amiaux}, J., {Arduini}, S., {et~al.} 2011, ArXiv e-prints,
  astro-ph/1110.3193

\bibitem[{{Le F{\`e}vre} {et~al.}(2013){Le F{\`e}vre}, {Cassata}, {Cucciati},
  {Garilli}, {Ilbert}, {Le Brun}, {Maccagni}, {Moreau}, {Scodeggio}, {Tresse},
  {Zamorani}, {Adami}, {Arnouts}, {Bardelli}, {Bolzonella}, {Bondi},
  {Bongiorno}, {Bottini}, {Cappi}, {Charlot}, {Ciliegi}, {Contini}, {de la
  Torre}, {Foucaud}, {Franzetti}, {Gavignaud}, {Guzzo}, {Iovino}, {Lemaux},
  {L{\'o}pez-Sanjuan}, {McCracken}, {Marano}, {Marinoni}, {Mazure}, {Mellier},
  {Merighi}, {Merluzzi}, {Paltani}, {Pell{\`o}}, {Pollo}, {Pozzetti},
  {Scaramella}, {Tasca}, {Vergani}, {Vettolani}, {Zanichelli}, \&
  {Zucca}}]{VVDS}
{Le F{\`e}vre}, O., {Cassata}, P., {Cucciati}, O., {et~al.} 2013, \aap, 559,
  A14

\bibitem[{{Lonsdale} {et~al.}(2003){Lonsdale}, {Smith}, {Rowan-Robinson},
  {Surace}, {Shupe}, {Xu}, {Oliver}, {Padgett}, {Fang}, {Conrow},
  {Franceschini}, {Gautier}, {Griffin}, {Hacking}, {Masci}, {Morrison},
  {O'Linger}, {Owen}, {P{\'e}rez-Fournon}, {Pierre}, {Puetter}, {Stacey},
  {Castro}, {Polletta}, {Farrah}, {Jarrett}, {Frayer}, {Siana}, {Babbedge},
  {Dye}, {Fox}, {Gonzalez-Solares}, {Salaman}, {Berta}, {Condon}, {Dole}, \&
  {Serjeant}}]{Lonsdale2003}
{Lonsdale}, C.~J., {Smith}, H.~E., {Rowan-Robinson}, M., {et~al.} 2003, \pasp,
  115, 897

\bibitem[{{Magnier} \& {Cuillandre}(2004)}]{Magnier2004}
{Magnier}, E.~A. \& {Cuillandre}, J.-C. 2004, \pasp, 116, 449

\bibitem[{{M{\'e}nard} {et~al.}(2003){M{\'e}nard}, {Hamana}, {Bartelmann}, \&
  {Yoshida}}]{Menard2003}
{M{\'e}nard}, B., {Hamana}, T., {Bartelmann}, M., \& {Yoshida}, N. 2003, \aap,
  403, 817

\bibitem[{{Milkeraitis} {et~al.}(2010){Milkeraitis}, {van Waerbeke}, {Heymans},
  {Hildebrandt}, {Dietrich}, \& {Erben}}]{Milkeraitis2010}
{Milkeraitis}, M., {van Waerbeke}, L., {Heymans}, C., {et~al.} 2010, \mnras,
  406, 673

\bibitem[{{Morrison} {et~al.}(2012){Morrison}, {Scranton}, {M{\'e}nard},
  {Schmidt}, {Tyson}, {Ryan}, {Choi}, \& {Wittman}}]{Morrison2012}
{Morrison}, C.~B., {Scranton}, R., {M{\'e}nard}, B., {et~al.} 2012, \mnras,
  426, 2489

\bibitem[{{Munari} {et~al.}(2013){Munari}, {Biviano}, {Borgani}, {Murante}, \&
  {Fabjan}}]{Munari2013}
{Munari}, E., {Biviano}, A., {Borgani}, S., {Murante}, G., \& {Fabjan}, D.
  2013, \mnras, 430, 2638

\bibitem[{{Muzzin} {et~al.}(2008){Muzzin}, {Wilson}, {Lacy}, {Yee}, \&
  {Stanford}}]{Muzzin2008}
{Muzzin}, A., {Wilson}, G., {Lacy}, M., {Yee}, H.~K.~C., \& {Stanford}, S.~A.
  2008, \apj, 686, 966

\bibitem[{{Muzzin} {et~al.}(2012){Muzzin}, {Wilson}, {Yee}, {Gilbank},
  {Hoekstra}, {Demarco}, {Balogh}, {van Dokkum}, {Franx}, {Ellingson}, {Hicks},
  {Nantais}, {Noble}, {Lacy}, {Lidman}, {Rettura}, {Surace}, \&
  {Webb}}]{Muzzin2012}
{Muzzin}, A., {Wilson}, G., {Yee}, H.~K.~C., {et~al.} 2012, \apj, 746, 188

\bibitem[{Muzzin {et~al.}(2009)Muzzin, Wilson, Yee, Hoekstra, Gilbank, Surace,
  Lacy, Blindert, Majumdar, Demarco, Gardner, Gladders, \&
  Lonsdale}]{Muzzin2009}
Muzzin, A., Wilson, G., Yee, H. K.~C., {et~al.} 2009, The Astrophysical
  Journal, 698, 1934

\bibitem[{{Muzzin} {et~al.}(2007){Muzzin}, {Yee}, {Hall}, \&
  {Lin}}]{Muzzin2007b}
{Muzzin}, A., {Yee}, H.~K.~C., {Hall}, P.~B., \& {Lin}, H. 2007, \apj, 663, 150

\bibitem[{{Narayan}(1989)}]{Narayan1989}
{Narayan}, R. 1989, \apjl, 339, L53

\bibitem[{{Navarro} {et~al.}(1997){Navarro}, {Frenk}, \& {White}}]{Navarro1997}
{Navarro}, J.~F., {Frenk}, C.~S., \& {White}, S.~D.~M. 1997, \apj, 490, 493

\bibitem[{{Planck Collaboration} {et~al.}(2016){Planck Collaboration}, {Ade},
  {Aghanim}, {Arnaud}, {Ashdown}, {Aumont}, {Baccigalupi}, {Banday},
  {Barreiro}, {Bartlett}, \& et~al.}]{Planck2015}
{Planck Collaboration}, {Ade}, P.~A.~R., {Aghanim}, N., {et~al.} 2016, \aap,
  594, A13

\bibitem[{{Raichoor} {et~al.}(2014){Raichoor}, {Mei}, {Erben}, {Hildebrandt},
  {Huertas-Company}, {Ilbert}, {Licitra}, {Ball}, {Boissier}, {Boselli},
  {Chen}, {C{\^o}t{\'e}}, {Cuillandre}, {Duc}, {Durrell}, {Ferrarese},
  {Guhathakurta}, {Gwyn}, {Kavelaars}, {Lan{\c c}on}, {Liu}, {MacArthur},
  {Muller}, {Mu{\~n}oz}, {Peng}, {Puzia}, {Sawicki}, {Toloba}, {Van Waerbeke},
  {Woods}, \& {Zhang}}]{Raichoor2014}
{Raichoor}, A., {Mei}, S., {Erben}, T., {et~al.} 2014, \apj, 797, 102

\bibitem[{{Rieke} {et~al.}(2004){Rieke}, {Young}, {Engelbracht}, {Kelly},
  {Low}, {Haller}, {Beeman}, {Gordon}, {Stansberry}, {Misselt}, {Cadien},
  {Morrison}, {Rivlis}, {Latter}, {Noriega-Crespo}, {Padgett}, {Stapelfeldt},
  {Hines}, {Egami}, {Muzerolle}, {Alonso-Herrero}, {Blaylock}, {Dole}, {Hinz},
  {Le Floc'h}, {Papovich}, {P{\'e}rez-Gonz{\'a}lez}, {Smith}, {Su}, {Bennett},
  {Frayer}, {Henderson}, {Lu}, {Masci}, {Pesenson}, {Rebull}, {Rho}, {Keene},
  {Stolovy}, {Wachter}, {Wheaton}, {Werner}, \& {Richards}}]{Rieke2004}
{Rieke}, G.~H., {Young}, E.~T., {Engelbracht}, C.~W., {et~al.} 2004, \apjs,
  154, 25

\bibitem[{{Rozo} \& {Schmidt}(2010)}]{Rozo2010}
{Rozo}, E. \& {Schmidt}, F. 2010, ArXiv e-prints, astro-ph/1009.5735

\bibitem[{{Schrabback} {et~al.}(2016){Schrabback}, {Applegate}, {Dietrich},
  {Hoekstra}, {Bocquet}, {Gonzalez}, {von der Linden}, {McDonald}, {Morrison},
  {Raihan}, {Allen}, {Bayliss}, {Benson}, {Bleem}, {Chiu}, {Desai}, {Foley},
  {de Haan}, {High}, {Hilbert}, {Mantz}, {Massey}, {Mohr}, {Reichardt}, {Saro},
  {Simon}, {Stern}, {Stubbs}, \& {Zenteno}}]{Schrabback2017}
{Schrabback}, T., {Applegate}, D., {Dietrich}, J.~P., {et~al.} 2016, ArXiv
  e-prints, astro-ph/1611.03866

\bibitem[{{Scranton} {et~al.}(2005){Scranton}, {M{\'e}nard}, {Richards},
  {Nichol}, {Myers}, {Jain}, {Gray}, {Bartelmann}, {Brunner}, {Connolly},
  {Gunn}, {Sheth}, {Bahcall}, {Brinkman}, {Loveday}, {Schneider}, {Thakar}, \&
  {York}}]{Scranton2005}
{Scranton}, R., {M{\'e}nard}, B., {Richards}, G.~T., {et~al.} 2005, \apj, 633,
  589

\bibitem[{{Seljak} \& {Warren}(2004)}]{Seljak2004}
{Seljak}, U. \& {Warren}, M.~S. 2004, \mnras, 355, 129

\bibitem[{{Skrutskie} {et~al.}(2006){Skrutskie}, {Cutri}, {Stiening},
  {Weinberg}, {Schneider}, {Carpenter}, {Beichman}, {Capps}, {Chester},
  {Elias}, {Huchra}, {Liebert}, {Lonsdale}, {Monet}, {Price}, {Seitzer},
  {Jarrett}, {Kirkpatrick}, {Gizis}, {Howard}, {Evans}, {Fowler}, {Fullmer},
  {Hurt}, {Light}, {Kopan}, {Marsh}, {McCallon}, {Tam}, {Van Dyk}, \&
  {Wheelock}}]{Skrutskie2006}
{Skrutskie}, M.~F., {Cutri}, R.~M., {Stiening}, R., {et~al.} 2006, \aj, 131,
  1163

\bibitem[{{Steidel} {et~al.}(1998){Steidel}, {Adelberger}, {Dickinson},
  {Giavalisco}, {Pettini}, \& {Kellogg}}]{Steidel1998}
{Steidel}, C.~C., {Adelberger}, K.~L., {Dickinson}, M., {et~al.} 1998, \apj,
  492, 428

\bibitem[{{Steidel} {et~al.}(1996){Steidel}, {Giavalisco}, {Pettini},
  {Dickinson}, \& {Adelberger}}]{Steidel1996}
{Steidel}, C.~C., {Giavalisco}, M., {Pettini}, M., {Dickinson}, M., \&
  {Adelberger}, K.~L. 1996, \apjl, 462, L17

\bibitem[{{Takada}(2010)}]{SHSC}
{Takada}, M. 2010, in American Institute of Physics Conference Series, Vol.
  1279, American Institute of Physics Conference Series, ed. N.~{Kawai} \&
  S.~{Nagataki}, 120--127

\bibitem[{{The Dark Energy Survey Collaboration}(2005)}]{DES}
{The Dark Energy Survey Collaboration}. 2005, ArXiv Astrophysics e-prints,
  astro-ph/0510346

\bibitem[{Umetsu {et~al.}(2011)Umetsu, Broadhurst, Zitrin, Medezinski, \&
  Hsu}]{Umetsu2011a}
Umetsu, K., Broadhurst, T., Zitrin, A., Medezinski, E., \& Hsu, L.-Y. 2011, The
  Astrophysical Journal, 729, 127

\bibitem[{{Umetsu} {et~al.}(2014){Umetsu}, {Medezinski}, {Nonino}, {Merten},
  {Postman}, {Meneghetti}, {Donahue}, {Czakon}, {Molino}, {Seitz}, {Gruen},
  {Lemze}, {Balestra}, {Ben{\'{\i}}tez}, {Biviano}, {Broadhurst}, {Ford},
  {Grillo}, {Koekemoer}, {Melchior}, {Mercurio}, {Moustakas}, {Rosati}, \&
  {Zitrin}}]{Umetsu2014}
{Umetsu}, K., {Medezinski}, E., {Nonino}, M., {et~al.} 2014, \apj, 795, 163

\bibitem[{{van der Burg} {et~al.}(2010){van der Burg}, {Hildebrandt}, \&
  {Erben}}]{vanderBurg2010}
{van der Burg}, R.~F.~J., {Hildebrandt}, H., \& {Erben}, T. 2010, \aap, 523,
  A74

\bibitem[{{van der Burg} {et~al.}(2014){van der Burg}, {Muzzin}, {Hoekstra},
  {Wilson}, {Lidman}, \& {Yee}}]{vanderBurg2014}
{van der Burg}, R.~F.~J., {Muzzin}, A., {Hoekstra}, H., {et~al.} 2014, \aap,
  561, A79

\bibitem[{{Van Waerbeke} {et~al.}(2010){Van Waerbeke}, {Hildebrandt}, {Ford},
  \& {Milkeraitis}}]{vanWaerbeke2010b}
{Van Waerbeke}, L., {Hildebrandt}, H., {Ford}, J., \& {Milkeraitis}, M. 2010,
  \apjl, 723, L13

\bibitem[{{Vandame}(2001)}]{Vandame2001}
{Vandame}, B. 2001, in Mining the Sky, ed. A.~J. {Banday}, S.~{Zaroubi}, \&
  M.~{Bartelmann}, 595

\bibitem[{{Voit}(2005)}]{Voit2005}
{Voit}, G.~M. 2005, Reviews of Modern Physics, 77, 207

\bibitem[{{von der Linden} {et~al.}(2014{\natexlab{a}}){von der Linden},
  {Allen}, {Applegate}, {Kelly}, {Allen}, {Ebeling}, {Burchat}, {Burke},
  {Donovan}, {Morris}, {Blandford}, {Erben}, \& {Mantz}}]{vonderLinden2014a}
{von der Linden}, A., {Allen}, M.~T., {Applegate}, D.~E., {et~al.}
  2014{\natexlab{a}}, \mnras, 439, 2

\bibitem[{{von der Linden} {et~al.}(2014{\natexlab{b}}){von der Linden},
  {Mantz}, {Allen}, {Applegate}, {Kelly}, {Morris}, {Wright}, {Allen},
  {Burchat}, {Burke}, {Donovan}, \& {Ebeling}}]{vonderLinden2014b}
{von der Linden}, A., {Mantz}, A., {Allen}, S.~W., {et~al.} 2014{\natexlab{b}},
  \mnras, 443, 1973

\bibitem[{{Wilson} {et~al.}(2005){Wilson}, {Gladders}, {Hoekstra}, {Lacy},
  {Muzzin}, {Surace}, \& {Yee}}]{Wilson2005b}
{Wilson}, G., {Gladders}, M., {Hoekstra}, H., {et~al.} 2005, {Detecting
  Clusters of Galaxies at 1 < z < 2 in the SWIRE Legacy Fields.}, Spitzer
  Proposal

\bibitem[{Wilson {et~al.}(2009)Wilson, Muzzin, Yee, Lacy, Surace, Gilbank,
  Blindert, Hoekstra, Majumdar, Demarco, Gardner, Gladders, \&
  Lonsdale}]{Wilson2009}
Wilson, G., Muzzin, A., Yee, H. K.~C., {et~al.} 2009, The Astrophysical
  Journal, 698, 1943

\bibitem[{{Wright} \& {Brainerd}(2000)}]{Wright2000}
{Wright}, C.~O. \& {Brainerd}, T.~G. 2000, \apj, 534, 34

\bibitem[{{Yee} \& {Ellingson}(2003)}]{Yee2003}
{Yee}, H.~K.~C. \& {Ellingson}, E. 2003, \apj, 585, 215

\bibitem[{{Yee} {et~al.}(1996){Yee}, {Ellingson}, \& {Carlberg}}]{Yee1996}
{Yee}, H.~K.~C., {Ellingson}, E., \& {Carlberg}, R.~G. 1996, \apjs, 102, 269

\bibitem[{{Yee} \& {L{\'o}pez-Cruz}(1999)}]{Yee1999}
{Yee}, H.~K.~C. \& {L{\'o}pez-Cruz}, O. 1999, \aj, 117, 1985

\end{thebibliography}
\clearpage

\end{document}